\documentclass[prd,a4paper,showpacs]{revtex4}
\usepackage{epsfig}
\usepackage{color}
\usepackage{amssymb}
\usepackage{amsfonts}
\usepackage{graphicx}
\usepackage{keyval,graphicx}
\usepackage{textcomp,wasysym}
\usepackage[dvipdfm, pdfstartview=FitH]{hyperref}
\newcommand{\ud}{\mathrm{d}}

\begin{document}

\title{Understanding the property of $\eta(1405/1475)$ in the $J/\psi$ radiative decay}

\author{Xiao-Gang Wu$^{1}$}
\author{Jia-Jun Wu$^{1,2}$}
\author{Qiang Zhao$^{1}$\footnote{zhaoq@ihep.ac.cn}}
\author{Bing-Song Zou$^{1,3}$\footnote{zoubs@ihep.ac.cn} }

\affiliation{ 1) Theoretical Physics Center for Science Facilities,
Institute of High Energy Physics, Chinese Academy of Sciences,
Beijing 100049, China \\
2) Physics Division, Argonne National Laboratory, Argonne, Illinois
60439, USA \\
3) State Key Laboratory of Theoretical Physics, Institute of
Theoretical Physics, Chinese Academy of Sciences, Beijing 100190,
China }

\begin{abstract}

In this work we make a systematic analysis of the correlated
processes $J/\psi\to \gamma \eta(1440)/f_1(1420)$ with
$\eta(1440)/f_1(1420)\to K\bar{K}\pi$, $\eta\pi\pi$ and $3\pi$,
where the role played by the so-called ``triangle singularity
mechanism" (TSM) is clarified. Our results agree well with the
experimental data and suggest a small fraction of $f_1(1420)$
contributions in these processes. This study confirms our conclusion
in [Phys. Rev. Lett. 108, 081803 (2012)] that the dynamic feature of
the TSM can be recognized by the strong narrow peak observed in the
$\pi\pi$ invariant mass spectrum of $\eta(1440)\to 3\pi$ with
anomalously large isospin violations. Nevertheless, we explicitly
demonstrate that the TSM can produce obvious peak position shifts
for the same $\eta(1440)$ or $f_1(1420)$ state in different decay
channels. This is a strong evidence that the $\eta(1405)$ and
$\eta(1475)$ are actually the same state, i.e. $\eta(1440)$. We also
make an analysis of the radiative decays of $\eta(1440)\to \gamma V$
($V=\phi$, $\rho^0$ or $\omega$) which shows that such a one-state
prescription seems not to have a conflict with the so-far existing
experimental data. Our analysis may shed a light on the
long-standing puzzling question on the nature of $\eta(1405)$ and
$\eta(1475)$.

\end{abstract}

\date{\today}





\maketitle

\section{Introduction}

The charmonium hadronic and radiative decays into light hadrons have
provided an important way to probe the light hadron structures. In
particular, with high statistics of $J/\psi$ and $\psi'$ events
produced in $e^+e^-$ annihilation, the light hadron spectra can be
studied closely and dynamic information concerning the light hadron
properties can be extracted from their production and decays. During
the past few years, there have been several new resonance structures
with $J^{PC}=0^{-+}$ observed by BESII and BESIII in $J/\psi$ and
$\psi'$ decays. They could be candidates of radial excitation states
of the pseudoscalar mesons $\eta$ and $\eta'$, or exotic states such
as glueball, multiquark state or hadronic molecule. For instance,
the BES-II Collaboration first reported a resonance structure in
$J/\psi\to \gamma X(1835) \to \gamma\eta^\prime
\pi^+\pi^-$~~\cite{Ablikim:2005um}, which was later confirmed by the
BESIII measurement~\cite{Ablikim:2010au} with high statistics.
Nevertheless, two additional resonance structures were identified as
$X(2120)$ and $X(2370)$ in the $\eta'\pi\pi$ invariant mass
spectrum~\cite{Ablikim:2010au, Yu:2011ta}.

In fact, our understanding of the isoscalar spectrum is still far
from well-established. Historically, the study of the nature of
$\eta(1405)$ and $\eta(1475)$ has been a hot topic and closely
related to the effort of searching for the ground state pseudoscalar
glueball in experiment. Since the first radial excitation states of
$\eta$ and $\eta'$ are generally assigned to $\eta(1295)$ and
$\eta(1475)$ taking into account their production and decay
properties~\cite{Nakamura:2010zzi}, it leaves out the abundant
$\eta(1405)$ as a possible candidate for the pseudoscalar glueball.
However, we would like to emphasize that such an arrangement still
needs further studies, and it is still controversial whether
$\eta(1405)$ and $\eta(1475)$ are two separated states or just one
state of $0^{-+}$ in different decay modes~\cite{Klempt:2007cp}.

With the availability of high-statistic $J/\psi$ and $\psi^\prime$
events from the BESIII Collaboration, it allows us to  tackle the
question on the nature of $\eta(1405)$ and $\eta(1475)$. One
important experimental progress is that the BESIII
Collaboration~\cite{BESIII:2012aa} report the observation of
anomalously large isospin violations of the $\eta(1405/1475)\to
3\pi$ in $J/\psi\to \gamma\eta(1405/1475) \to \gamma \pi^0
f_0(980)\to \gamma +3\pi$ which, however, can hardly be understood
by treating them as either glueball or $q\bar{q}$ state.
Interestingly, this decay process also explicitly involves the issue
of $a_0(980)$-$f_0(980)$ mixings. The BESIII data show that the
$f_0(980)$ signal is only about 10 MeV in width and the lineshape is
different from the Breit-Wigner width of about $40\sim 100$
MeV~\cite{Nakamura:2010zzi}. Moreover, the isospin violation turns
out to be significant with $BR(\eta(1405)\to f_0(980)\pi^0\to
3\pi)/BR(\eta(1405)\to a^0_0(980)\pi^0\to \eta\pi\pi)=(17.9\pm
4.2)\%$, which cannot be explained by the $a_0-f_0$ mixing intensity
measured in other channels~\cite{Ablikim:2010aa}.

An immediate theoretical interpretation is given by
Ref.~\cite{Wu:2011yx}, where we propose that a triangle singularity
mechanism (TSM) via the intermediate $K^*\bar{K}+c.c.$ rescatterings
would lead to significant enhancement of the isospin violating
decay, i.e. $\eta(1405/1475)\to K^*\bar{K}+c.c.\to f_0(980)\pi$. In
this transition, the dominant contributions would come from such a
specific kinematic region that all the intermediate mesons in the
triangle loop are literally on-shell. The identification of such a
mechanism seems to be nontrivial since it can naturally explain the
narrow width of the $f_0(980)$ observed in the two pion invariant
mass spectrum. Consequently, it raises an essential issue concerning
the nature of $\eta(1405/1475)$ since the TSM can also contribute to
the decays of $\eta(1405/1475)\to K\bar{K}\pi$ and $\eta\pi\pi$, and
distort the lineshapes and shift the peak positions of the
$\eta(1405/1475)$ in those decay channels. As a result, a coherent
study of $\eta(1405/1475)\to K\bar{K}\pi$, $\eta\pi\pi$, and $3\pi$
is necessary and could be a key towards a better understanding of
the $\eta(1405/1475)$ puzzle.

In this work, we shall provide a detailed analysis of
$\eta(1405/1475)\to K\bar{K}\pi$, $\eta\pi\pi$, and $3\pi$. We shall
show that only one $0^{-+}$ isoscalar state, namely $\eta(1440)$, is
needed in this mass region. With this ``one state" assumption, we
shall demonstrate that the TSM can lead to different mass spectra
for $\eta(1440)\to K\bar{K}^*+c.c.$, $a_0(980)\pi^0$, and
$f_0(980)\pi$. In $J/\psi\to \gamma\eta(1440)$, with $\eta(1440)\to
K\bar{K}^*+c.c.$, $a_0(980)\pi^0$, and $f_0(980)\pi$, another
possible contribution to the same final states is via $f_1(1420)$.
Since the mass of $f_1(1420)$ is similar to that of $\eta(1440)$, we
should investigate the role played by $f_1(1420)$ in these
processes. In particular, due to the similar masses between
$f_1(1420)$ and $\eta(1440)$, the decay of $f_1(1420)$ would also
experience the TSM. Therefore, a helicity analysis of the invariant
mass spectrum for the overlapping $f_1(1420)$ and $\eta(1440)$ is
necessary. In comparison with the results reported in
Ref.~\cite{Wu:2011yx}, we have detailed all the analysis by
including the $f_1(1420)$ contributions. We confirm the BESIII
results by detailed helicity analysis from which we can extract the
invariant mass spectra for $\eta(1440)$ in different channels. These
features as a consequence of the TSM could be a natural solution for
the long-standing puzzle about the nature of $\eta(1405/1475)$ in
experimental analyses.

We also mention that the $\eta(1405/1475)\to 3\pi$ decay was also
studied in Ref.~\cite{Aceti:2012dj} recently in a chiral unitary
approach.  By exhausting several models and taking constraints from
the meson-meson scatterings, the authors confirm that only
$a_0(980)$-$f_0(980)$ mixing can not explain the BES result
\cite{BESIII:2012aa} and the inclusion of the triangular diagrams is
necessary~\cite{Wu:2011yx}.

The rest part of this paper is organized as follows: the formalism
is presented in Sec. II. Section III is devoted to the numerical
results and discussions. Our conclusion is given in Sec. IV.

\section{Formalism}

\subsection{Effective Lagrangians and transition amplitudes}

\begin{figure}[htbp]
  \centering
  \vspace{-1cm}
  \includegraphics[scale=0.8]{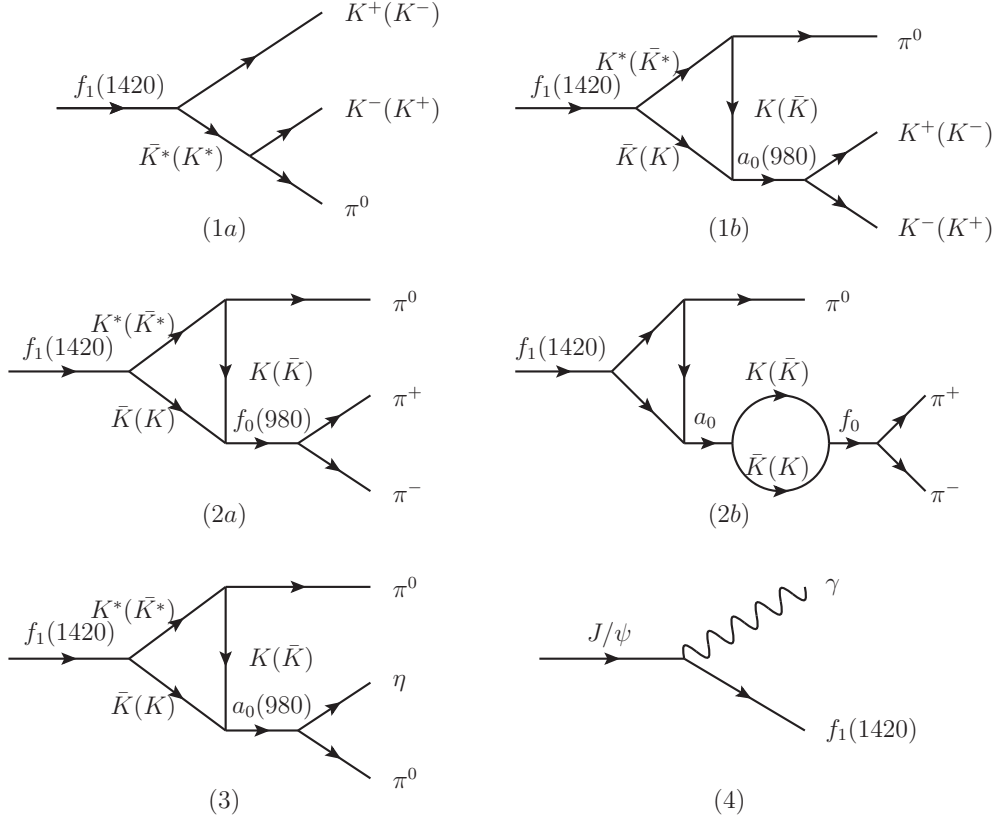}\\
  \vspace{-11cm}
\caption{Feynman diagrams for the $f_1(1420)$ decays and its
production in $J/\psi$ radiative decay. Similar diagrams for
$\eta(1440)$ have been given in Ref.~\cite{Wu:2011yx}. }\label{fig1}
\end{figure}

The effective Lagrangians for the $\eta(1440)$ production have been
presented in Ref.~\cite{Wu:2011yx}. Here, we include the $f_1(1420)$
contribution and list the effective Lagrangians as the following:
\begin{eqnarray}
  \mathcal{L}_{V_1 V_2 P} &=& g_{V_1 V_2 P} \varepsilon_{\mu \nu \rho \sigma} p_{V_1}^{\mu} p_{V_2}^{\nu} \psi_{V_1}^{\rho} \psi_{V_2}^{\sigma} \psi_P \ , \label{a1:1}\\
  \mathcal{L}_{V P_1 P_2} &=& g_{V P_1 P_2} ( \psi_{P1} \partial_{\mu} \psi_{P_2} - \psi_{P2} \partial_{\mu} \psi_{P_1} ) \psi_V^{\mu} \ ,  \label{a1:2}\\
  \mathcal{L}_{S P_1 P_2} &=& g_{S P_1 P_2} \psi_{S} \psi_{P_1} \psi_{P_2} \ ,  \label{a1:3}\\
  \mathcal{L}_{AVP} &=& g_{AVP} \psi_{A}^{\mu} \psi_{V \mu} \psi_P \ ,  \label{a1:4}\\
  \mathcal{L}_{\psi \gamma f_1} &=& g_1 \varepsilon_{\mu \nu \rho \sigma} \partial^\mu \psi_{\psi}^{\nu} \psi_{\gamma}^{\rho} \psi_{f_1}^{\sigma} + g_2 \varepsilon_{\mu \nu \rho \sigma} \partial^{\mu} \psi_{\psi}^{\lambda} \partial^\lambda \partial^\nu \psi_{\gamma}^{\rho} \psi_{f_1}^{\sigma} \ , \label{a1:5}
\end{eqnarray}
where $S$, $P$, $V$ and $A$ stand for four types of fields: scalar,
pseudoscalar, vector and axialvector, respectively. For $\eta(1440)$
the same diagrams as in Ref.~\cite{Wu:2011yx} are calculated, while
for $f_1(1420)$ the similar diagrams are listed in Fig.~\ref{fig1}.
Figure~\ref{fig1}(1a)-(1b) are for $f_1(1420)\to K\bar{K}\pi$
through $K^*\bar{K}$ and $a_0(980)\pi$ channels.
Figure~\ref{fig1}(2a)-(2b) are for $f_1(1420)\to \pi^+ \pi^- \pi^0$
through the TSM and $a_0 - f_0$ mixing. Figure~\ref{fig1}(3) is for
$f_1(1420) \to \eta \pi^0 \pi^0$, where we assume that
$a_0^0(980)\pi^0$ gives the main contribution.

\begin{figure}
  \centering
  \includegraphics[scale=0.8]{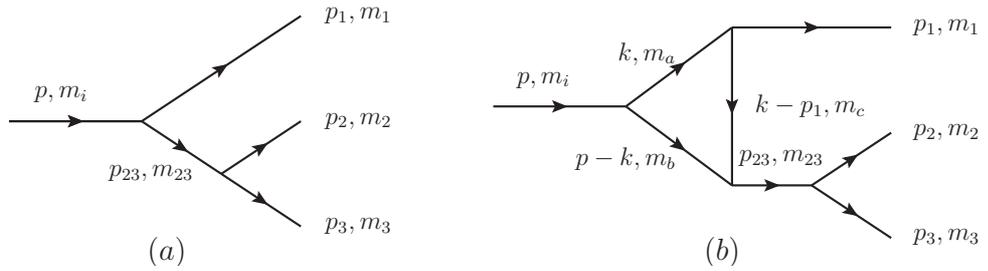}\\
  \vspace{-18cm}
  \caption{Kinematics defined in our formalism.}\label{fig2}
\end{figure}

Our kinematics conventions are shown in Fig.~\ref{fig2}. Some common
functions are defined as follows:
\begin{eqnarray}
  G_f&=& \frac{1}{s - m_f^2 + i \sqrt{s} \Gamma_f(s)} \ , \label{a3:1}\\
  G_a&=& \frac{1}{s - m_a^2 + i \sqrt{s} \Gamma_a(s)} \ , \label{a3:2}\\
  \Gamma_{a}(s)&=&\frac{g^2_{aK\bar{K}}(\rho(\sqrt{s},m_{K^0},m_{\bar{K}^0})
  +\rho(\sqrt{s},m_{K^+},m_{K^-}))}{16\pi\sqrt{s}}+\frac{g^2_{a\pi\eta}\rho(\sqrt{s},m_{\pi^0},m_{\eta})}{16\pi\sqrt{s}} \ , \label{a3:3}\\
  \Gamma_{f}(s)&=&\frac{g^2_{fK\bar{K}}(\rho(\sqrt{s},m_{K^0},m_{\bar{K}^0})+\rho(\sqrt{s},m_{K^+},m_{K^-}))}{16\pi\sqrt{s}}\nonumber\\
  &&+\frac{g^2_{f\pi\pi}(\rho(\sqrt{s},m_{\pi^0},m_{\pi^0})+2\rho(\sqrt{s},m_{\pi^+},m_{\pi^-}))}{16\pi\sqrt{s}} \ ,  \label{a3:4}\\
  \rho(\sqrt{s},m_A,m_B)&=&\frac{1}{s}\sqrt{(s-(m_A+m_B)^2)(s-(m_A-m_B)^2)} \ . \label{a3:5}
\end{eqnarray}
Then, the typical loop integrals can be expressed as
\begin{eqnarray}
  \hat{I}_{\eta1} &=& i \int \frac{\ud^4 k}{(2\pi)^4} (2p-k)^\mu (2p_1-k)^\nu \frac{(-g_{\mu \nu}+\frac{k_\mu k_\nu}{m_a^2})}{k^2-m_a^2}
  \frac{1}{(p-k)^2-m_b^2} \frac{1}{(k-p_1)^2-m_c^2} \ ,\label{a3:6}\\
  \hat{I}_{f1} &=&
  i \int \frac{\ud^4 k}{(2\pi)^4} \epsilon^\mu_p (2p_1-k)^\nu
  \frac{(-g_{\mu \nu}+\frac{k_\mu k_\nu}{m_a^2})}{k^2-m_a^2}
  \frac{1}{(p-k)^2-m_b^2} \frac{1}{(k-p_1)^2-m_c^2}
  \nonumber  \\
  &=& \epsilon_p^\mu (c_p p_\mu + c_{p_1} p_{1\mu}) \ ,
  \label{a3:7}\\
  \hat{I}_{f1b} &=&
  \hat{I}_{f1}(1 \leftrightarrow 2) = \epsilon_p^\mu ( d_p p_\mu + d_{p_2}p_{2\mu} ) \ .  \label{a3:8}
\end{eqnarray}
Taking into account that the relative signs between the charged and
neutral loops are positive in isospin-conserving processes but
negative in isospin-violating processes, it is convenient to define
\begin{eqnarray}
  c_p^+ &\equiv & c_p^c g_{k^* K \pi}^c + c_p^n g_{k^* K \pi}^n \ ,
  \label{a4:1}\\
  c_p^- &\equiv & c_p^c g_{k^* K \pi}^c - c_p^n g_{k^* K \pi}^n \ ,
  \label{a4:2}\\
  c_{p_1}^+ &\equiv & c_{p_1}^c g_{k^* K \pi}^c + c_{p_1}^n g_{k^* K \pi}^n \ ,
  \label{a4:3}\\
  c_{p_1}^- &\equiv & c_{p_1}^c g_{k^* K \pi}^c - c_{p_1}^n g_{k^* K \pi}^n \ ,
  \label{a4:4}
\end{eqnarray}
where the superscripts `c' and `n' denote the charged and neutral
loops, respectively. We also define $\hat{I}_{\eta1}^{\pm}$,
$d_p^{\pm}$ and $d_{p_2}^{\pm}$ for $\hat{I}_{f1b}$ in a similar
way.

The invariant amplitudes in Fig.~\ref{fig1} can then be expressed as
\begin{eqnarray}
  \label{1a}
  \mathcal{M}_{1a} &=&
  g_{f_1 K^* K} g_{K^* K \pi}\left[\epsilon_{p}^{\mu} \frac{(-g_{\mu \nu} + p_{23\mu} p_{23\nu})}{s_{23}-m_V^2 + i m_V \Gamma_V} (p_3-p_2)^\nu + (2 \leftrightarrow 1) \right]
  \nonumber \\
  &=&
  g_{f_1 K^* K} g_{K^* K \pi} \times \epsilon_p^\mu \left( c_1 p_{1\mu} + c_2 p_{2\mu} + c_3 p_{3\mu} \right) \ ,
  \\
  \label{1a2}
  \overline{\sum_{spin}} |\mathcal{M}_{1a}|^2 &=&
  \frac{1}{3} g_{f_1 K^* K}^2 g_{K^* K \pi}^2 \left( -g^{\mu \nu} + \frac{p^\mu p^\nu}{m_{f_1}^2} \right) \left( c_1 p_{1\mu} + c_2 p_{2\mu} + c_3 p_{3\mu} \right)  \left( c_1^* p_{1\nu} + c_2^* p_{2\nu} + c_3^* p_{3\nu} \right) \ ,
  \\
  \label{1b}
  \mathcal{M}_{1b} &=&
  2 g_{f_1 K^* K} g_{a K \bar{K}}^2 G_a(s_{23}) \left( g_{K^* K \pi}^c \hat{I}_{f1}^c + g_{K^* K \pi}^n \hat{I}_{f1}^n \right)
  \nonumber \\
  &=&
  2 g_{f_1 K^* K} g_{a K \bar{K}}^2 G_a(s_{23}) \times \epsilon_p^\mu \left( p_\mu c_p^+ + p_{1\mu} c_{p_1}^+ \right) \ ,
  \\
  \label{1b2}
  \overline{\sum_{spin}} |\mathcal{M}_{1b}|^2 &=&
  \frac{4}{3} g_{f_1 K^* K}^2 g_{aK\bar{K}}^4 |G_a(s_{23})|^2 \left( -g^{\mu \nu} + \frac{p^\mu p^\nu}{m_{f_1}^2} \right) (p_\mu c_p^+ + p_{1\mu} c_{p_1}^+) (p_\nu c_p^{+*} + p_{1\nu} c_{p_1}^{+*}) \ ,
  \\
  \label{2a}
  \mathcal{M}_{2a} &=&
  2 \sqrt{2} g_{f_1 K^* K} g_{fK\bar{K}} g_{f\pi\pi} G_f(s_{23}) \left( g_{K^*K\pi}^c \hat{I}_{f1}^c - g_{K^*K\pi}^n \hat{I}_{f1}^n \right)
  \nonumber \\
  &=&
  2 \sqrt{2} g_{f_1 K^* K} g_{fK\bar{K}} g_{f\pi\pi} G_f(s_{23}) \times \epsilon_p^\mu \left( p_\mu c_p^- + p_{1\mu} c_{p_1}^- \right) \ ,
  \\
  \label{2a2}
  \overline{\sum_{spin}} |\mathcal{M}_{2a}|^2 &=&
  \frac{8}{3} g_{f_1 K^* K}^2 g_{fK\bar{K}}^2 g_{f\pi\pi}^2 |G_f(s_{23})|^2 \left( -g^{\mu \nu} + \frac{p^\mu p^\nu}{m_{f_1}^2} \right) (p_\mu c_p^- + p_{1\mu} c_{p_1}^-) (p_\nu c_p^{-*} + p_{1\nu} c_{p_1}^{-*}) \ ,
  \\
  \label{2b}
  \mathcal{M}_{2b} &=& \mathcal{M}_{1b} \times \sqrt{2} g_{fK\bar{K}} g_{f\pi\pi} G_f(s_{23})\left( loop2^c -loop2^n \right) \ ,
  \\
  \label{2b2}
  \overline{\sum_{spin}} |\mathcal{M}_{2b}|^2 &=& \left(\overline{\sum_{spin}} |\mathcal{M}_{1b}|^2\right) \times 2 g_{fK\bar{K}}^2 g_{f\pi\pi}^2 |G_f(s_{23})|^2 |loop2^c-loop2^n|^2 \ ,
  \\
  \label{3a}
  \mathcal{M}_{3} &=&
  2 g_{f_1 K^* K} g_{aK\bar{K}} g_{a \pi \eta} \left[ G_a(s_{23}) \left( g_{K^* K \pi}^c \hat{I}_{f1}^c + g_{K^* K \pi}^n \hat{I}_{f1}^n \right) + (2 \leftrightarrow 1) \right]
  \nonumber \\
  &=&
  2 g_{f_1 K^* K} g_{aK\bar{K}} g_{a \pi \eta} \times \epsilon_p^\mu \left[ G_a(s_{23}) \left( p_\mu c_p^+ + p_{1\mu} c_{p_1}^+ \right) + (2 \leftrightarrow 1) \right] \ ,
  \\
  \label{3a2}
  \overline{\sum_{spin}} |\mathcal{M}_{3}|^2 &=&
  \frac{2}{3} g_{f_1 K^* K}^2 g_{aK\bar{K}}^2 g_{a \pi \eta}^2 \left( -g^{\mu \nu} + \frac{p^\mu p^\nu}{m_{f_1}^2} \right)
  [
  |G_a(s_{23})|^2 (p_\mu c_p^+ + p_{1\mu} c_{p_1}^+)(p_\nu c_p^{+*} + p_{1\nu} c_{p_1}^{+*})
  \nonumber \\&&
  + |G_a(s_{13})|^2 (p_\mu d_p^+ + p_{2\mu} d_{p_2}^+)(p_\nu d_p^{+*} + p_{2\nu} d_{p_2}^{+*})
  \nonumber \\&&
  + G_a(s_{23}) G_a(s_{13})^* (p_\mu c_p^+ + p_{1\mu} c_{p_1}^+)(p_\nu d_p^{+*} + p_{2\nu} d_{p_2}^{+*})
  \nonumber \\&&
  + G_a(s_{23})^* G_a(s_{13}) (p_\mu c_p^{+*} + p_{1\mu} c_{p_1}^{+*})(p_\nu d_p^+ + p_{2\nu} d_{p_2}^+)
  ] \ ,
  \\
  \label{4a}
  \mathcal{M}_4 &=& g_1 \epsilon_{\mu\nu\rho\sigma} p_{\psi}^{\mu} \epsilon_{\psi}^{\nu} \epsilon_{\gamma}^{\rho} \epsilon_{f_1}^{\sigma}
  + g_2 \epsilon_{\mu\nu\rho\sigma} p_{\psi}^{\mu} p_{f_1}^{\nu} \epsilon_{\gamma}^{\rho} \epsilon_{f_1}^{\sigma}
  \epsilon_{\psi} \cdot p_{f_1} B_2(Q) \ .
\end{eqnarray}
In Eq.~(\ref{3a2}), we have put the identical factor $1/2$ in the
squared amplitude. The parametrization of Eq.~(\ref{4a}) is taken
from Ref.~\cite{Zou:2002ar} and  $B_2(Q)$ is the Blatt-Weisskopf
barrier factor
\begin{equation}\label{a5}
    B_2(Q)=\sqrt{\frac{1}{Q^4 + 3Q^2 Q_0^2 + 9Q_0^4}} \ ,
\end{equation}
where $Q$ is the decay momentum,  and $Q_0$ is a hadron scale
parameter $Q_0=0.197321/R \ \textrm{GeV}$ with $R$ the radius of the
centrifugal barrier in fermi. In this paper we adopt $R=0.35 \
\textrm{fm}$ which is about the radius of $J/\psi$.

\subsubsection{Helicity amplitudes}

In experiment, the quantum number of an intermediate state $X$ is
generally determined by measuring the angular distribution of the
$X$ decays. To proceed, we first make a model-independent analysis
of the helicity structure of the transition matrix element which
would allow us to separate different partial waves. Then, by
comparing with the angular distributions measured by experiment, we
can extract the dynamic coupling strengths for different partial
waves.

For a decay process $a \to b+c$ with spin, helicity and parity
$(s_i,\lambda_i,\eta_i)_{i=a,b,c}$, the decay amplitude in the rest
frame of $a$ can be expressed as~\cite{Chung:1997jn},
\begin{equation}\label{angular:1}
    \mathcal{M}_{\lambda_b \lambda_c}^{s_a}(\theta,\phi;\lambda_a)\propto
    D_{\lambda_a, \lambda_b-\lambda_c}^{s_a *}\!(\phi, \theta, 0) F_{\lambda_b
    \lambda_c}^{s_a} \ ,
\end{equation}
where $D_{\lambda_a, \lambda_b-\lambda_c}^{s_a *}\!(\phi, \theta,
0)$ is the rotation function, and $F_{\lambda_b \lambda_c}^{s_a}$ is
the helicity-coupling amplitude which is independent of angular
variables. It satisfies two constraints taking into account angular
momentum conservation and Parity conversation:
\begin{eqnarray}\label{angular:2}
  |\lambda_b -\lambda_c| &\leq& s_a \ ,\\
  F_{\lambda_b \lambda_c}^{s_a} &=& \eta_a \eta_b \eta_c (-)^{s_a - s_b - s_c} F_{-\lambda_b
  -\lambda_c}^{s_a} \ .
\end{eqnarray}
From these relations, we can find out the independent
helicity-coupling amplitudes.

As follows, we will re-analyze the angular distributions of the
recoiled photon in $J/\psi\to \gamma X$ and the recoiled $f_0(980)$
in $X\to f_0(980)\pi$ decays which are measured by the BESIII
experiment~\cite{BESIII:2012aa}.
Now as a preparation, we derive the helicity amplitudes of
each vertex from the Lagrangians in Eqs.~(\ref{a1:1})-(\ref{a1:5}).

In $J/\psi \to \gamma f_1$, the helicity amplitude with transversely
polarized $f_1$ can be expressed as
  \begin{eqnarray}\label{h1:psi_photon_f1}
    \langle \lambda_\gamma, \lambda_{f_1}=\pm 1| \hat{S} | \lambda_\psi \rangle &=&
    g_1 \epsilon_{\mu \nu \rho \sigma} p_{\psi}^{\mu} \epsilon_{\psi}^{\nu} \epsilon_{\gamma}^{*\rho} \epsilon_{f_1}^{*\sigma} +
    g_2 B_2(Q) \epsilon_{\mu \nu \rho \sigma} p_{\psi}^{\mu} p_\gamma^{\nu} \epsilon_{\gamma}^{*\rho} \epsilon_{f_1}^{*\sigma}
    \epsilon_{\psi} \cdot p_\gamma  \nonumber\\
     &=& -i g_1 \lambda_\gamma m_{\psi} D_{\lambda_\psi 0}^{1*}(\phi, \theta, 0) + i g_2 B_2(Q) \lambda_\gamma m_{\psi} Q^2 D_{\lambda_\psi 0}^{1*}(\phi, \theta, 0) \nonumber\\
     &=& D_{\lambda_\psi 0}^{1*}(\phi, \theta, 0) F_{\lambda_\gamma
     \lambda_{f_1}}^{1} \ ,
  \end{eqnarray}
which is nonvanishing with $\lambda_\gamma =\lambda_{f_1}$, and
\begin{equation}\label{h2:psi_photon_f1}
    F_{\lambda_\gamma \lambda_{f_1}}^{1}= -i g_1 \lambda_\gamma m_\psi + i g_2 \lambda_\gamma m_\psi Q^2
    B_2(Q) \ .
\end{equation}
So the independent amplitude is
  \begin{equation}\label{h3:psi_photon_f1}
    F_{11}^{1a}=i m_\psi[-g_1+g_2 Q^2 B_2(Q)] \ .
  \end{equation}
When $f_1$ is longitudinally polarized, the $g_2$ term will have no
contribution. The helicity amplitude is
  \begin{eqnarray}\label{h4:psi_photon_f1}
    \langle \lambda_\gamma, \lambda_{f_1}=0| \hat{S} | \lambda_\psi \rangle  &=&
    g_1 \epsilon_{\mu \nu \rho \sigma} p_{\psi}^{\mu} \epsilon_{\psi}^{\nu} \epsilon_{\gamma}^{*\rho} \epsilon_{f_1}^{*\sigma}  \nonumber\\
&=& -i g_1 \lambda_\gamma m_\psi \frac{E_{f_1}}{m_{f_1}}
D_{\lambda_\psi \lambda_\gamma}^{1*}(\phi, \theta, 0) \nonumber\\
    &=& D_{\lambda_\psi \lambda_\gamma}^{1*}(\phi, \theta, 0) F_{\lambda_\gamma
    0}^{1} \ ,
  \end{eqnarray}
  with
  \begin{equation}\label{h5:psi_photon_f1}
    F_{\lambda_\gamma 0}^{1} = -i g_1 \lambda_\gamma m_\psi
    \frac{E_{f_1}}{m_{f_1}} \ .
  \end{equation}
So the independent amplitude is
  \begin{equation}\label{h6:psi_photon_f1}
    F_{10}^{1a} = -i g_1 m_\psi \frac{E_{f_1}}{m_{f_1}} \ .
  \end{equation}

For $f_1 \to f_0(980) \pi^0$, the helicity amplitude can be written
as
  \begin{eqnarray}\label{h1: f1_f0_pi}
    \langle f_0 \pi | \hat{S} | f_1 \rangle &=&
    2 g_{f_1 K^* K} g_{f_0 K K} \left[ g_{K^* K \pi}^c \hat{I}_{f1}^c - g_{K^* K \pi}^n \hat{I}_{f1}^n \right) \nonumber\\
    &=&
    2 g_{f_1 K^* K} g_{f_0 K K} \epsilon_{p}^{\mu} ( p_{\mu} c_p^- + p_{1\mu} c_{p_1}^-) \nonumber\\
    &=&
    2 g_{f_1 K^* K} g_{f_0 K K} c_{p_1}^- \epsilon_{p} \cdot p_1 \nonumber\\
    &=& D_{\lambda_{f_1} 0}^{1*}(\phi, \theta, 0) F_{00}^{1b} \ ,
  \end{eqnarray}
where we have used the relation $\epsilon_{p} \cdot p =0$. Also, in
the helicity frame of $f_1(1420)$, we have
  \begin{equation}\label{h2: f1_f0_pi}
    \epsilon_{p} \cdot p_1 = - D_{\lambda_f 0}^{1*}(\phi, \theta, 0) Q
    \ .
  \end{equation}
So the independent amplitude is
\begin{equation}\label{h3: f1_f0_pi}
    F_{00}^{1b} = -2 g_{f_1 K^* K} g_{f_0 K K} c_{p_1}^- Q \ .
\end{equation}

Similarly, the helicity amplitude for $J/\psi \to \gamma \eta(1440)$
can be obtained with both $J/\psi$ and $\gamma$ transversely
polarized:
  \begin{eqnarray}\label{h1:psi_photon_eta1}
    \langle \lambda_\gamma \eta(1440)| \hat{S} | \lambda_\psi \rangle &=&
    g_{\psi \gamma \eta_1} \epsilon_{\mu \nu \rho \sigma} p_{\psi}^{\mu} q_{\gamma}^{\nu} \epsilon_{\psi}^{\rho} \epsilon_{\gamma}^{\sigma}
    \nonumber\\
    &=&
    -i \lambda_\gamma g_{\psi \gamma \eta_1} m_{\psi} |\vec{q}|
D_{\lambda_\psi \lambda_\gamma}^{1*}(\phi, \theta, 0) \nonumber\\
    &=& D_{\lambda_\psi \lambda_\gamma}^{1*}(\phi, \theta, 0) F_{\lambda_\gamma
    0}^{1c} \ ,
  \end{eqnarray}
with
  \begin{eqnarray}\label{h2:psi_photon_eta1}
    F_{\lambda_\gamma 0}^{1c}  &=& -i \lambda_\gamma g_{\psi \gamma \eta_1} m_{\psi}
    |\vec{q}| \ .
  \end{eqnarray}

For $\eta(1440) \to f_0(980) \pi^0$, the helicity amplitude is
  \begin{eqnarray}\label{h1:eta1_f0_pi}
  &&  \langle f_0 \pi | \hat{S} | \eta(1440) \rangle \nonumber\\
  &=& 2 g_{\eta_1 K^* K} g_{f_0 K K} ( g_{K^* K \pi}^c \hat{I}_{f1}^c - g_{K^* K \pi}^n \hat{I}_{f1}^n ) \nonumber\\
  &=& F_{00}^{0d} \ .
  \end{eqnarray}

\subsection{Angular distribution}

\begin{figure}[htbp]
  \centering
  \vspace{-1cm}
  \includegraphics[scale=0.5]{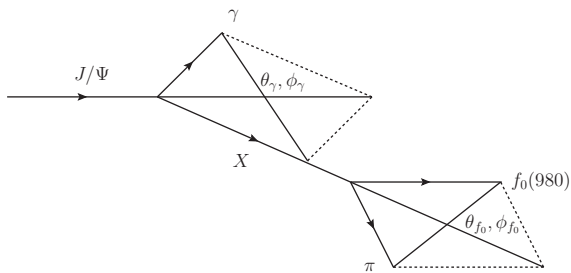}\\
  \vspace{-10cm}
  \caption{The kinematics for  the chain process
$J/\psi\to \gamma X \to \gamma f_0(980)\pi^0$. }\label{fig3}
\end{figure}

By combining the two-body decay amplitudes in the helicity frame, we
can derive the total helicity amplitudes for the chain process
$J/\psi\to \gamma X \to \gamma f_0(980)\pi^0$ as shown in Fig.~\ref{fig3}
and extract the angular distributions to
compare with the experimental data~\cite{BESIII:2012aa}.

For $X$ being $\eta(1440)$, the total helicity amplitude can  be
expressed as
\begin{equation}\label{angular:14}
    A_{\eta_1}(\lambda_\psi,\lambda_\gamma) \propto
    D_{\lambda_\psi \lambda_\gamma}^{1*}(\phi_\gamma, \theta_\gamma, 0) F_{\lambda_\gamma 0}^{1c}
    \frac{1}{s- m_{\eta_1}^2 + i m_{\eta_1} \Gamma_{\eta_1}}
    F_{00}^{0d} \ .
\end{equation}
The angular distribution is
\begin{eqnarray}\label{angular:15}
  \frac{\ud \sigma_{\eta_1}}{\ud \Omega} &\propto&
  \sum_{\lambda_\psi,\lambda_\gamma =\pm 1} \left| A_{\eta_1}(\lambda_\psi,\lambda_\gamma) \right|^2 \nonumber\\
  &\propto&
  \sum_{\lambda_\psi,\lambda_\gamma =\pm 1} \left| D_{\lambda_\psi \lambda_\gamma}^{1*}(\phi_\gamma, \theta_\gamma, 0) \right|^2 \nonumber\\
  &=& 1 + \cos^2 \theta_\gamma \ ,
\end{eqnarray}
where $\ud \Omega\equiv \ud \Omega_{\gamma} \ud \Omega_{f_0}$ with $\ud
\Omega_{\gamma}\equiv \ud \cos\theta_\gamma \ud \phi_\gamma$ and $\ud
\Omega_{f_0} \equiv \ud \cos\theta_{f_0} \ud \phi_{f_0}$. One step
further, we obtain
\begin{eqnarray}
  \frac{\ud \sigma_{\eta_1}}{\ud \cos\theta_\gamma} &\propto& 1 + \cos^2 \theta_\gamma \ ,\label{angular:16}\\
  \frac{\ud \sigma_{\eta_1}}{\ud \cos\theta_{f_0}} &\propto&
  \mathrm{const} \ , \label{angular-eta}
\end{eqnarray}
for the angular distributions of $\theta_\gamma$ and $\theta_{f_0(980)}$,
respectively. These expressions are the same as those adopted in
Ref.~\cite{BESIII:2012aa}.

For $X$ being $f_1(1420)$, the total helicity amplitude of the chain
process can be expressed
as
\begin{equation}\label{angular:7}
    A_{f_1}(\lambda_\psi, \lambda_\gamma, \lambda_{f_1}) \propto
    D_{\lambda_\psi,\lambda_\gamma -\lambda_{f_1}}^{1*}(\phi_\gamma, \theta_\gamma, 0) F_{\lambda_\gamma \lambda_{f_1}}^{1a}
    \frac{1}{s-m_{f_1}^2+i m_{f_1} \Gamma_{f_1}}
    D_{\lambda_{f_1} 0}^{1*}(\phi_{f_0}, \theta_{f_0}, 0) F_{00}^{1b}
    \ .
\end{equation}
The angular distribution via $f_1(1420)$ is
\begin{eqnarray}\label{angular:8}
\frac{\ud \sigma_{f_1}}{\ud \Omega} &\propto&
\sum_{\lambda_\psi,\lambda_\gamma =\pm 1} \left|
\sum_{\lambda_{f_1}=0,\pm1} A_{f_1}(\lambda_\psi, \lambda_\gamma,
\lambda_{f_1}) \right|^2  \nonumber\\
  &\propto& \sum_{\lambda_\psi,\lambda_\gamma =\pm 1} \left|
  \alpha D_{\lambda_\psi, 0}^{1*}(\phi_\gamma, \theta_\gamma, 0) D_{\lambda_\gamma 0}^{1*}(\phi_{f_0}, \theta_{f_0}, 0) +
  D_{\lambda_\psi, \lambda_\gamma}^{1*}(\phi_\gamma, \theta_\gamma, 0) D_{0 0}^{1*}(\phi_{f_0}, \theta_{f_0}, 0) \right|^2
  \nonumber\\
  &=& \alpha_1^2 \sin^2 \theta_{f_0} \sin^2 \theta_\gamma + \frac{\alpha_1}{2} \cos\phi_{\alpha} \cos\phi_{f_0} \sin2\theta_{f_0} \sin2\theta_\gamma + \cos^2 \theta_{f_0} (\cos^2 \theta_\gamma +
  1) \ ,
\end{eqnarray}
where  $\alpha\equiv \alpha_1 e^{i \phi_{\alpha}}$ is the
ratio of the $\lambda_{f_1}=\pm 1$ amplitude to that of
$\lambda_{f_1}=0$.
By integrating over corresponding polar angles in the above double
distribution, one has access to the angular distributions of
$\theta_\gamma$ and $\theta_{f_0}$, respectively, for the
intermediate $f_1(1420)$:
\begin{eqnarray}
  \frac{\ud \sigma_{f_1}}{\ud \cos\theta_\gamma}
  &\propto&
   1 + 2\alpha_1^2 + (1-2\alpha_1^2) \cos^2\theta_\gamma  \ ,\label{angular:10}\\
  \frac{\ud \sigma_{f_1}}{\ud \cos\theta_{f_0}}
  &\propto&
  2 + (\alpha_1^2 - 2) \sin^2\theta_{f_0} \ . \label{angular-f1}
\end{eqnarray}

With both $\eta(1440)$ and $f_1(1420)$ contributing to
the chain process, the total helicity amplitude
can be obtained in a similar way, namely,
\begin{eqnarray}
  A_{\eta_1+f_1}(\lambda_\psi,\lambda_\gamma) &=&
  D_{\lambda_\psi \lambda_\gamma}^{1*}(\phi_\gamma, \theta_\gamma, 0)
  F_{\lambda_\gamma 0}^{1c}
  \frac{1}{s- m_{\eta_1}^2 + i m_{\eta_1} \Gamma_{\eta_1}}
  F_{00}^{0d} \nonumber\\
  && +
  D_{\lambda_\psi 0}^{1*}(\phi_\gamma, \theta_\gamma, 0)
  D_{\lambda_\gamma 0}^{1*}(\phi_{f_0}, \theta_{f_0}, 0)
  F_{\lambda_\gamma \lambda_\gamma}^{1a}
  \frac{1}{s-m_{f_1}^2+i m_{f_1} \Gamma_{f_1}}
  F_{00}^{1b}
    \nonumber\\
  && +
  D_{\lambda_\psi \lambda_\gamma}^{1*}(\phi_\gamma, \theta_\gamma, 0)
  D_{0 0}^{1*}(\phi_{f_0}, \theta_{f_0}, 0)
  F_{\lambda_\gamma 0}^{1a}
  \frac{1}{s-m_{f_1}^2+i m_{f_1} \Gamma_{f_1}}
  F_{00}^{1b}
   \nonumber\\
  &\propto& \lambda_\gamma \left[
    r D_{\lambda_\psi \lambda_\gamma}^{1*}(\phi_\gamma, \theta_\gamma, 0)
  + \alpha D_{\lambda_\psi 0}^{1*}(\phi_\gamma, \theta_\gamma, 0) D_{\lambda_\gamma 0}^{1*}(\phi_{f_0}, \theta_{f_0}, 0)
  + D_{\lambda_\psi \lambda_\gamma}^{1*}(\phi_\gamma, \theta_\gamma, 0)  D_{0 0}^{1*}(\phi_{f_0}, \theta_{f_0}, 0)
  \right] \ , \label{angular:17}
\end{eqnarray}
where we have applied the selection rule
$\lambda_{f_1}=\lambda_\gamma$ for the transversely polarized
$f_1(1420)$; $r\equiv r_1 e^{i \phi_r}$ is the ratio of the
$\eta(1440)$ amplitude to that of $f_1(1420)$ with
$\lambda_{f_1}=0$. Then the angular distribution becomes
\begin{eqnarray}\label{angular:18}
  \frac{\ud \sigma_{\eta_1+f_1}}{\ud \Omega} &\propto&
  \sum_{\lambda_\psi,\lambda_\gamma =\pm 1} \left|A_{\eta_1+f_1}(\lambda_\psi,\lambda_\gamma) \right|^2
  \nonumber\\
  &=&
  r_1^2 (\cos2\theta_\gamma + 3) + 2 r_1 \alpha_1 \sin\theta_{f_0} \sin2\theta_\gamma \cos\phi_{f_0} \cos(\phi_\alpha-\phi_r) +
  2 r_1 \cos\theta_{f_0} \cos\phi_r (\cos2\theta_r + 3)
  \nonumber\\&&
  + 2 \alpha_1^2 \sin^2\theta_{f_0} \sin^2\theta_\gamma +
  \alpha_1 \cos\phi_\alpha \sin2\theta_{f_0} \sin2\theta_\gamma \cos\phi_{f_0} + \cos^2\theta_{f_0} (\cos2\theta_\gamma +
  3) \ .
\end{eqnarray}
It is easy to show that Eqs.~(\ref{angular:15}) and
(\ref{angular:8}) can be reproduced by setting the corresponding
resonance couplings to vanish. Similarly, the angular distributions
of $\theta_\gamma$ and $\theta_{f_0}$ can be obtained
\begin{eqnarray}
  \frac{\ud \sigma_{\eta_1+f_1}}{\ud \cos\theta_\gamma} &\propto&
  1 + 2 \alpha_1^2 + 3 r_1^2 + (1 - 2\alpha_1^2 + 3 r_1^2)
  \cos^2\theta_\gamma \ ,\label{angular:19}
  \\
  \frac{\ud \sigma_{\eta_1+f_1}}{\ud \cos\theta_{f_0}} &\propto&
  \alpha_1^2 + 2 r_1^2 + 4 r_1 \cos\phi_r \cos\theta_{f_0} + (2 - \alpha_1^2) \cos^2
  \theta_{f_0} \ .\label{angular-both}
\end{eqnarray}
Checking Eqs.~(\ref{angular:16}), (\ref{angular:10}) and
(\ref{angular:19}), one can see that the $\cos\theta_\gamma$ distribution
is always symmetric as a feature of a two-body decay. In contrast,
the angular distribution of $\cos\theta_{f_0}$ turns out to be
nontrivial. As shown by Eqs.~(\ref{angular-eta}), (\ref{angular-f1})
and (\ref{angular-both}), the contributions from different states
with different quantum numbers are encoded in the angular
distribution of  $\cos\theta_{f_0}$. By fitting the experimental
data, the coupling parameters can thus be determined which
alternatively would provide information about the contributing
resonances. As shown by Fig.~3 of Ref.~\cite{BESIII:2012aa}, the
$\cos\theta_{f_0}$ distribution is apparently asymmetric which
indicates some contributions from the $f_1(1420)$ production besides
$\eta(1440)$.

With the explicit total helicity amplitude in
Eq.~(\ref{angular:17}), we can express the differential width as
\begin{equation}\label{heli:1}
    \ud\Gamma = \frac{1}{(2\pi)^5} \frac{1}{16 m_\psi^2}  \left| A_{\eta_1+f_1} \right|^2 |\vec{p}_\gamma| |\vec{p}_{f_0}|
    \ud \sqrt{s_X} \ud\Omega_\gamma \ud\Omega_{f_0} \ .
\end{equation}
We can define the following quantities by integrating the invariant mass
$\sqrt{s_X}$ of the $f_0(980)\pi^0$:
\begin{eqnarray}\label{heli:2}
  A_1 &=& \int\ud \sqrt{s_X} \; |\vec{p}_\gamma| |\vec{p}_{f_0}|
  \left|
  \frac{F_{1 0}^{1c} F_{00}^{0d}}{s_X- m_{\eta_1}^2 + i m_{\eta_1} \Gamma_{\eta_1}}
  \right|^2 \ ,
  \\
  A_2 &=& \int\ud \sqrt{s_X} \; |\vec{p}_\gamma| |\vec{p}_{f_0}|
  \left|
  \frac{F_{11}^{1a} F_{00}^{1b}}{s_X-m_{f_1}^2+i m_{f_1} \Gamma_{f_1}}
  \right|^2 \ ,
  \\
  A_3 &=& \int\ud \sqrt{s_X} \; |\vec{p}_\gamma| |\vec{p}_{f_0}|
  \left|
  \frac{F_{10}^{1a} F_{00}^{1b}}{s_X-m_{f_1}^2+i m_{f_1} \Gamma_{f_1}}
  \right|^2 \ ,
\end{eqnarray}
which can thus be related to the quantities measured in experiment,
i.e.
\begin{equation}\label{heli:3}
    \alpha_1^2=\frac{A_2}{A_3},\quad r_1^2=\frac{A_1}{A_3} \ .
\end{equation}
From these relations, we can extract the information about the
couplings from the angular distribution analysis.

To compare with the experimental measurement of the unpolarized
partial decay width in terms of the recoiled energy $s$ by the
photon in $J/\psi \to \gamma X \to \gamma ABC$, the following
standard expression is adopted,
\begin{equation}\label{a6}
\frac{\ud \Gamma_{J/\psi \to \gamma X \to \gamma ABC}}{\ud \sqrt{s}}
= \frac{2s}{\pi} \frac{\Gamma_{J/\psi \to \gamma X}(s) \times
\Gamma_{X \to ABC}(s)}{(s-m_X^2)^2 + \Gamma_{X}^2 m_X^2} \ ,
\end{equation}
where $s$ is the four-momentum square of $X=\eta(1440)/f_1(1420)$ in
the reaction.  A constant width $\Gamma(f_1(1420))=0.0549 \
\textrm{GeV}$ is adopted for $f_1(1420)$~\cite{Nakamura:2010zzi},
while for $\eta(1440)$, both constant width and energy-dependent
form are adopted,
\begin{equation}\label{a7}
    \Gamma_{\eta(1440)}(s)=\Gamma_{\eta(1440) \to K^* K \to
    K\bar{K}\pi}(s)=\Gamma_{1a}(s) \ ,
\end{equation}
where $\Gamma_{1a}$ corresponds to Fig.~\ref{fig1}(1a).

\section{Results and discussions}

In this part, we present our analyses and numerical results. First
we demonstrate explicitly that the TSM is dominant in $\eta(1440)\to
3\pi$, and the main contribution is indeed from such a kinematic
region that all the internal particles are close to their mass
shells. Then, we give the fitting results about $\eta(1440)$ from
$K\bar{K}\pi$ spectrum and show the predictions of $\pi^+\pi^-\pi^0$
and $\eta\pi^0\pi^0$ which are consistent with
Ref.~\cite{Wu:2011yx}. By including $f_1(1420)$, we extract the
couplings of $f_1(1420)$ through the analysis of the angular
distribution of $\pi^+\pi^-\pi^0$ channel~\cite{BESIII:2012aa}.
Finally, we show that the combined results for both $\eta(1440)$ and
$f_1(1420)$ in comparison with the BES data would allow us to draw a
conclusion on the anomalously large isospin violations observed in
$\eta(1405)\to 3\pi$ and the nature of $\eta(1405)$ and
$\eta(1475)$.

\subsection{Loop integral}

Here we discuss in detail the calculation of $\hat{I}_{\eta1}$ and
$\hat{I}_{f1}$ in Eqs.~(\ref{a3:6})-(\ref{a3:7}). What we actually
need in $\hat{I}_{f1}$ is the coefficients $c_p$ and $c_{p1}$. We
use two methods to calculate the loops:
\begin{enumerate}
\item We directly calculate $\hat{I}_{\eta1}$ and $\hat{I}_{f1}$ by {\it LoopTools} without any form factors.
The UV divergences are regularized dimensionally by
$\Delta=2/(4-D)-\gamma_E+\log{4\pi}$, where $\Delta$ can be adjusted
and the default value is $\Delta=0$.

\item The exponential form factor method as applied in Ref.~\cite{Wu:2011yb}.
Namely, an exponential form factor as follows is included in
$\hat{I}_{\eta1}$ and $\hat{I}_{f1}$ to cut off the UV divergence:
\begin{equation}\label{a8}
    \textrm{exp}\left[\frac{k^2-m_a^2}{\Lambda^2} + \frac{(p-k)^2-m_b^2}{\Lambda^2} + \frac{(k-p_1)^2-m_c^2}{\Lambda^2}\right] \ ,
\end{equation}
where $\Lambda$ is the cutoff energy and characterize the effective
range of the interaction. In principle, other forms of form factors
can also be examined and we find the results are similar to each
other.

\end{enumerate}
\begin{figure}[htbp]
  \centering
  \includegraphics[scale=1.6]{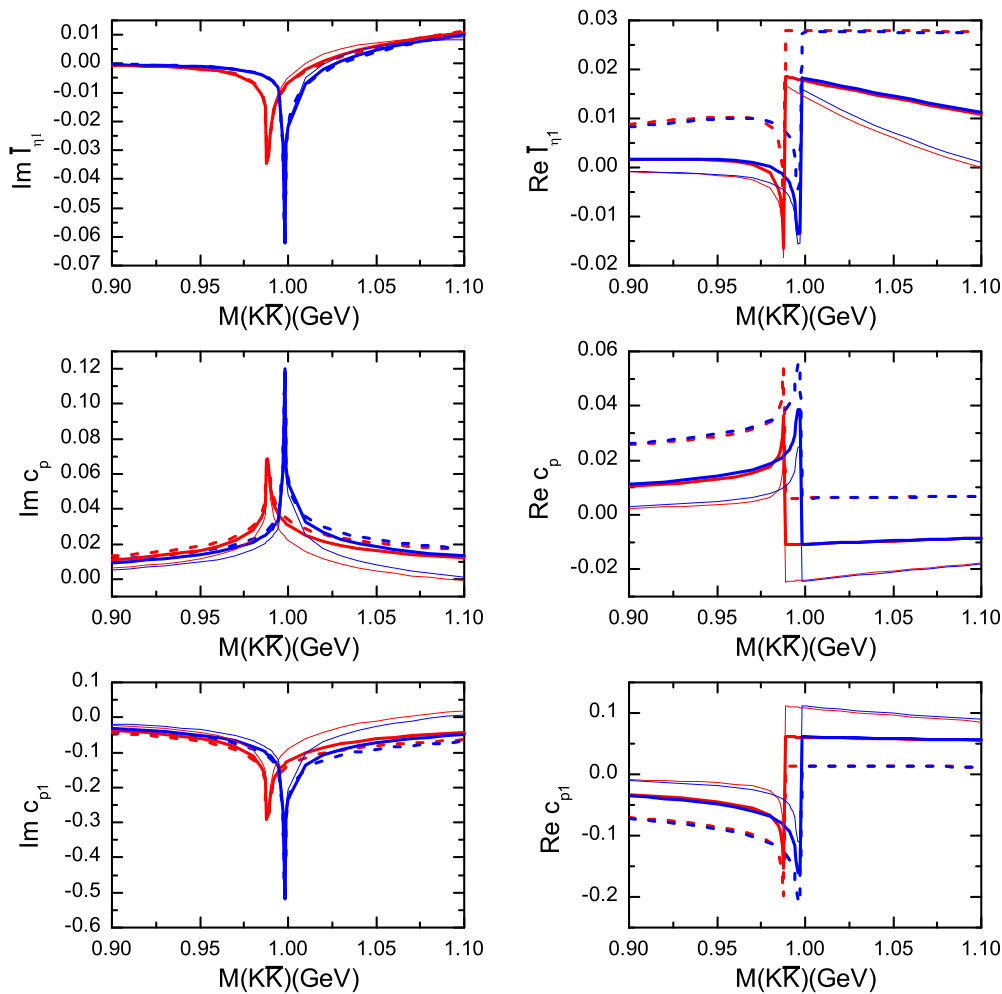}
  \vspace{-2cm}
\caption{The imaginary and real parts of the loop integrals in terms
of the invariant mass of $K\bar{K}$ at $\sqrt{s}=1.42 \
\textrm{GeV}$. There are two sets of lines identified by those two
spikes which correspond to the charged (lower mass) and neutral
(higher mass) $K\bar{K}$ thresholds. For each set of lines, the
thick and thin solid ones represent the results with $\Lambda=1.0 $
and $\Lambda=0.5 \ \textrm{GeV}$, respectively, while the dashed
lines denote the results of {\it LoopTools} calculation without form
factor.}
  \label{fig:loop3}
\end{figure}

We present the calculations based on the above two treatments in
Fig.~\ref{fig:loop3}. In the kinematic region that all the internal
particles are close to their mass shells, namely the TS kinematics,
these two treatments give nearly identical results since the form
factor corrections are nearly unity. In particular, the absorptive
part is dominated by the contributions from the TS kinematics. The
real part turns out to be more sensitive to the form factor
corrections when the internal particles deviate from their mass
shells. Similar results are found for the $f_1(1420)$ since its mass
is nearly the same as $\eta(1440)$ and they share the same TSM. As a
result, one can imagine that there should be little difference
between these two treatments in the isospin-violating decay of
$\eta(1440)/f_1(1420)\to 3\pi$ since the main contribution is from
the absorptive part in the TS kinematics and the dispersive part
would largely cancel out between the charged and neutral loop
amplitudes. It is worth noting that the cancellation between the
charged and neutral loop amplitudes eventually makes the calculation
almost independent of the model uncertainties as explicitly pointed
out in Ref.~\cite{Wu:2011yx}. This should be a direct way to confirm
the dominance of the TSM in $\eta\to 3\pi$ as a dynamic mechanism.

\subsection{Angular distribution analysis}

In the numerical calculations, the common coupling constants present
in the triangle loops for $\eta(1440)$ and $f_1(1420)$ are adopted
the same as in Ref.~\cite{Wu:2011yx}, i.e. $g_{aK\bar{K}}=3.33 \
\textrm{GeV}, \ g_{a \eta \pi}=2.45 \
\textrm{GeV},\,g_{K^*K\pi}^n=3.208$, and $g_{K^*K\pi}^c=3.268$. BES
~\cite{Ablikim:2004wn} and KLOE~\cite{Aloisio:2002bt} give different
values for the $f_0(980)$ coupling, namely, $g_{fK\bar{K}}=4.18 \
\textrm{GeV}$ and $g_{f\pi\pi}=1.66 \ \textrm{GeV}$ from
BES~\cite{Ablikim:2004wn}; and $g_{fK\bar{K}}=5.92 \ \textrm{GeV}$
and $g_{f\pi\pi}=2.09 \ \textrm{GeV}$ from
KLOE~\cite{Aloisio:2002bt}. Similar to the treatment in
Ref.~\cite{Wu:2011yx}, contributions from Figs.~\ref{fig1}(1b) and
(2b) are neglected since they are only about $1/10$ of
Figs.~\ref{fig1}(1a) and (2a), respectively.

We adopt the mass and width of $f_1(1420)$ from the
PDG~\cite{Nakamura:2010zzi}, i.e. $m_{f_1(1420)}=1.4264 \ \textrm{GeV}$
and $\Gamma_{f_1(1420)}=54.9 \ \textrm{MeV}$, but leave the mass and width of
$\eta(1440)$ to be fitted by the experimental data based on the
``one-state" assumption. This is reasonable since the $f_1$ spectrum
does not suffer from the ambiguity of possible abundant states in
this energy region and as we shall see later that the $f_0(980)$
angular distribution measured by BESIII~\cite{BESIII:2012aa} only
requires a small contribution from the $f_1(1420)$.

Taking into account the present datum status, our analysis strategy
is as follows: we first fit the BESIII data~\cite{BESIII:2012aa} for
the angular distributions of the recoiled photon and $f_0(980)$ in
the decay of $J/\psi\to \gamma X$ and $X\to f_0(980)\pi^0$. This
allows us to extract the relative strengths between the $\eta(1440)$
and $f_1(1420)$ in the isospin violating decays. Then, by applying
for the constraint from the $J/\psi\to \gamma\eta(1405/1475) \to
\gamma K\bar{K}\pi$ from DM2, MARK III, and
BES~\cite{Nakamura:2010zzi}, it allows us to determine the absolute
differential widths for both $\eta(1440)$ and $f_1(1420)$. We shall
compare this with the exclusive fit by $\eta(1440)$ as shown in
Ref.~\cite{Wu:2011yx}. In the end, we shall output the invariant
mass spectra for $\eta(1440)/f_1(1420)\to K\bar{K}\pi$, $\eta\pi\pi$
and $3\pi$, from which we would expect to observe different
lineshapes and peak positions from the same state in different decay
channels.

The $\theta_{\gamma}$ and $\theta_{f_0}$ angular distributions of
exclusive $\eta(1440)$ and $f_1(1420)$ have been analyzed in
Ref.~\cite{BESIII:2012aa} as parameterized in
Eqs~(\ref{angular:16})-(\ref{angular-eta}) and
(\ref{angular:10})-(\ref{angular-f1}). Now we consider the combined
contributions from both $\eta(1440)$ and $f_1(1420)$ and fit the
BESIII data~\cite{BESIII:2012aa} using
Eqs.~(\ref{angular:19})-(\ref{angular-both}) which can be expressed
as
\begin{eqnarray}\label{angular:20}
  \frac{\ud N}{\ud \cos\theta_\gamma} &=&
  b_\gamma(1 + c \cos^2\theta_\gamma) \ , \\
  \frac{\ud N}{\ud \cos\theta_{f_0}} &=&
  b_{f_0} (1 + c_1 \cos\theta_{f_0} + c_2 \cos^2\theta_{f_0}) \ ,
\end{eqnarray}
where $b_\gamma$ and $b_{f_0}$ are the overall normalization factors
and
\begin{equation}\label{angular:21}
  c   \equiv  \frac{1 - 2\alpha_1^2 + 3 r_1^2}{1 + 2 \alpha_1^2 + 3 r_1^2} ,\qquad
  c_1 \equiv  \frac{4 r_1 \cos\phi_r}{\alpha_1^2 + 2 r_1^2} ,\qquad
  c_2 \equiv  \frac{2 - \alpha_1^2}{\alpha_1^2 + 2 r_1^2} \ .
\end{equation}

\begin{figure}[htbp]
  \begin{minipage}[l]{0.45\textwidth}
  \centering
  \includegraphics[scale=0.3]{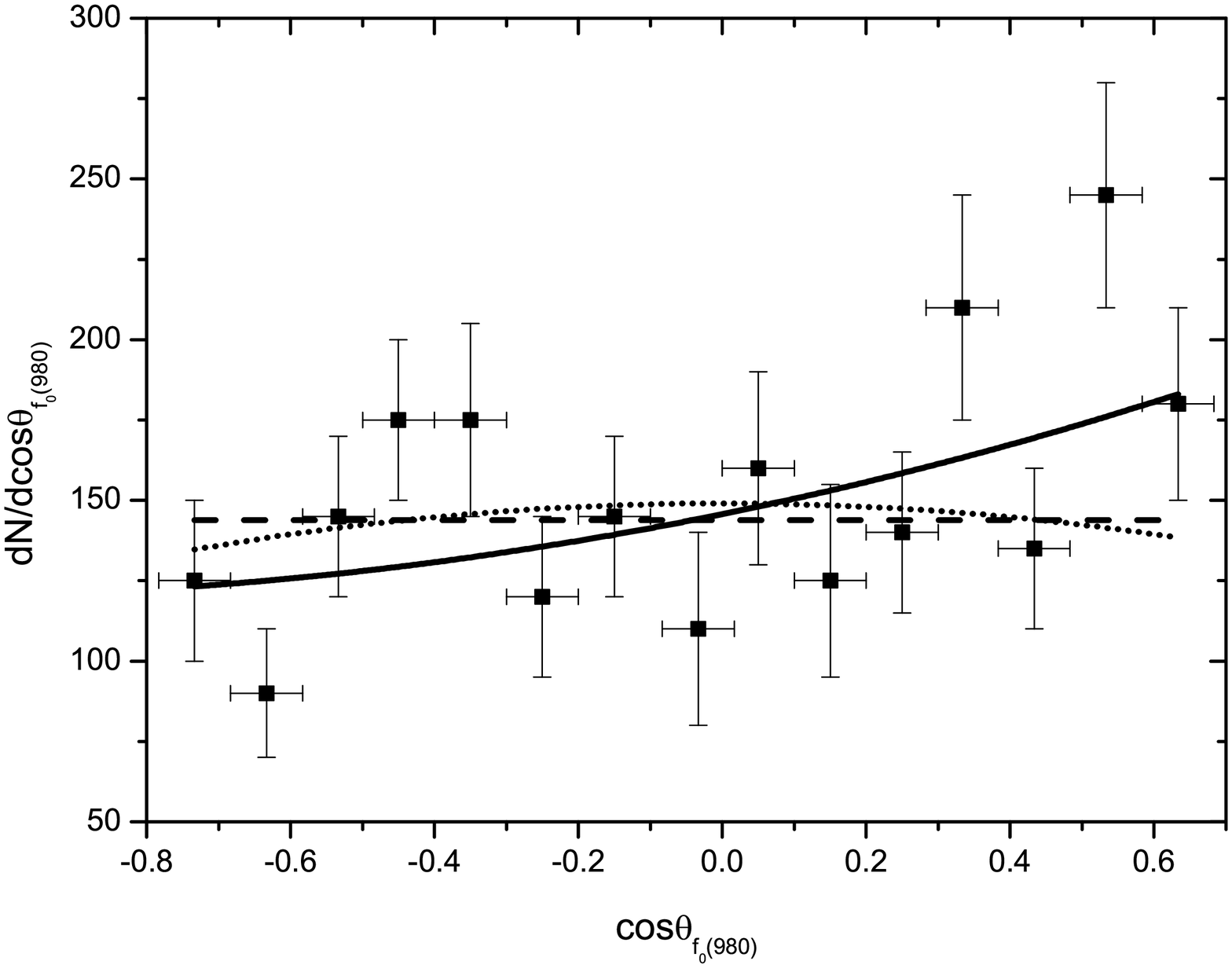}
  \end{minipage}
  \begin{minipage}[l]{0.45\textwidth}
  \centering
  \includegraphics[scale=0.3]{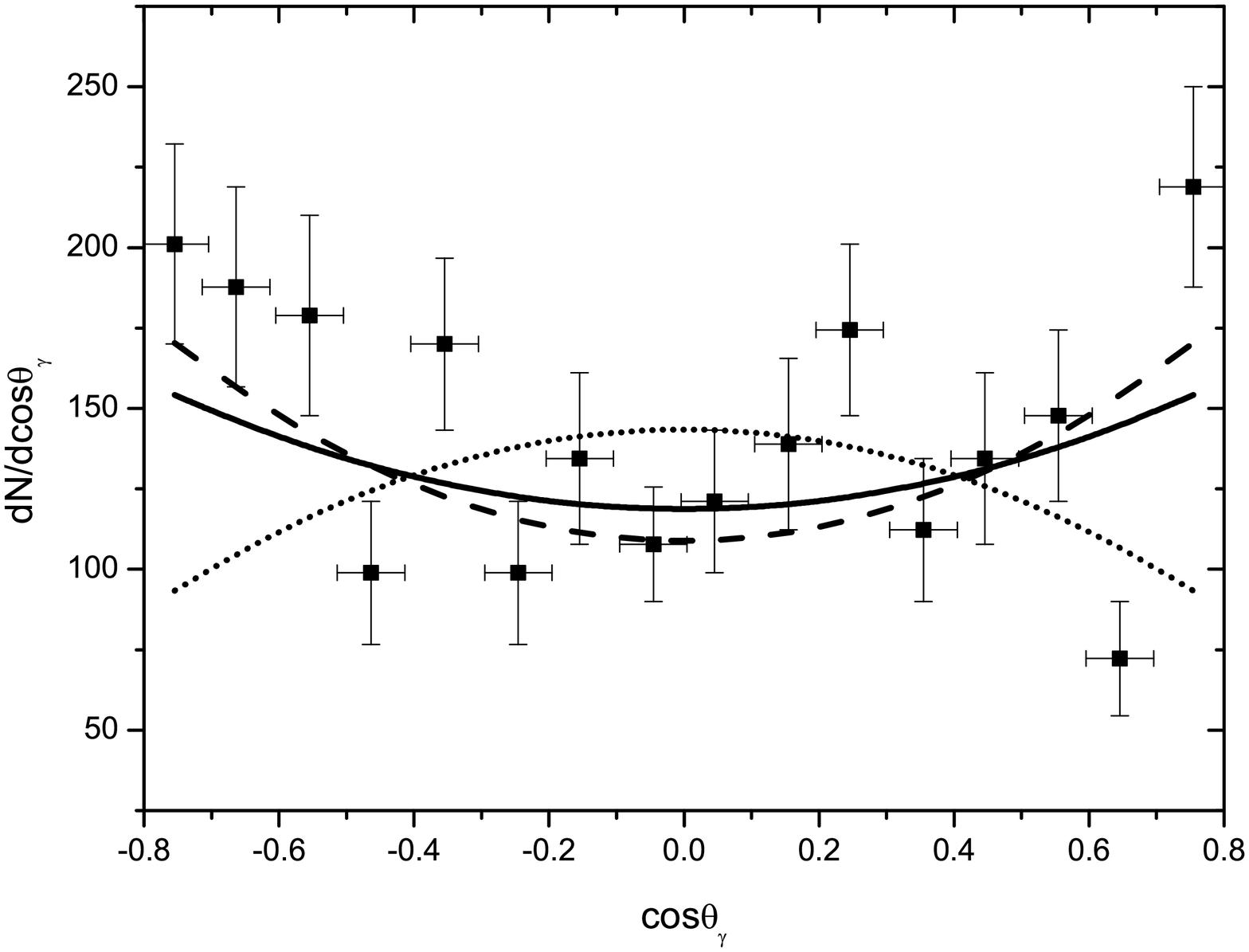}
  \end{minipage}
\caption{Fitting results for the $\cos\theta_{f_0}(980)$
distribution (left panel) and $\cos\theta_\gamma$ distribution
(right panel), respectively. The solid lines are the results
considering both $\eta(1440)$ and $f_1(1420)$, while the dashed and
dotted lines are the results with exclusive $\eta(1440)$ or
$f_1(1420)$, respectively.
  }
  \label{fig5-redefine}
\end{figure}

We use the CERN program MINUIT to fit the data and the fitting
results are demonstrated in Fig.~\ref{fig5-redefine}. To compare
with the results of Ref.~\cite{BESIII:2012aa},  we show the $\chi^2$
values of different fits in Table~\ref{tab:chisq}. From
Fig.~\ref{fig5-redefine} we can see that the angular distributions
are improved significantly when both $\eta(1440)$ and $f_1(1420)$
are included. The apparently asymmetric behavior of the
$\cos\theta_{f_0}$ distribution in Fig.~\ref{fig5-redefine} can be
well explained as the interference between $\eta(1440)$ and
$f_1(1420)$.

\begin{table}[htbp]
  \centering
  \caption{The fitting qualities of different fits.}\label{tab:chisq}
  \begin{tabular}{ccc}
    \hline\hline
    immediate states  &  $\chi^2/d.o.f$ for $\cos\theta_\gamma$ & $\chi^2/d.o.f$ for $\cos\theta_{f_0}$ \\ \hline
    $\eta(1440)$                 & $40.2/15$ &  $26.8/14$ \\
    $f_1(1420)$                  & $59.0/15$ &  $26.4/13$ \\
    $\eta(1440)$ and $f_1(1420)$ & $38.3/14$ &  $19.8/12$ \\
    \hline\hline
  \end{tabular}
\end{table}

The fitted parameters are as follows:
\begin{equation}\label{angular:22}
   b_\gamma=118.5 \pm 8.8, \ \ c=0.538 \pm 0.312 \ ,
\end{equation}
from the $\theta_\gamma$ distribution and
\begin{equation}\label{angular:23}
    b_{f_0}=145.7 \pm 10.7, \ \ c_1=0.314 \pm 0.128,\, c_2=0.141 \pm 0.317 \
    ,
\end{equation}
for $\theta_{f_0}$. By solving Eq.~(\ref{angular:21}), we obtain
\begin{equation}\label{angular:24}
    \alpha_1^2 = 1.197 \pm 1.090,\,
    r_1 = 1.50 \pm 0.89,\,
    \phi_r = \pm (1.27 \pm 0.20) \ ,
\end{equation}
which will allow us to extract the coupling constants for
$f_1(1420)$, i.e. $g_{f_1(1420)K^*K}$, $g_1$, and $g_2$. Coupling
$g_{f_1(1420)K^*K}$ can be directly obtained from the width of
$f_1(1420)$~\cite{Nakamura:2010zzi} by assuming that $K\bar{K}\pi$
channel is dominant; $g_{\psi \gamma \eta(1440)}/g_1$ is related to
the fitted $r_1$; and $g_2/g_1$ is related to the fitted $\alpha_1$.

The fitted parameters of $\eta(1440)$ are $m_{\eta(1440)}=1.42 \ \textrm{GeV}
\, , \, \Gamma_{\eta(1440)}=67\ \textrm{MeV}$.
The extracted couplings are listed in Table~\ref{tab:2}. There are
two solutions for the ratio of the $D$-wave coupling to $S$-wave
coupling $g_2/g_1$, i.e. the value $-0.179$ indicates the $S$-wave
dominant, while the value $0.970$ indicates the $D$-wave dominant.
The present precision of the experimental data seems impossible to
distinguish these two solutions. From our fit we find that the ratio
of $f_1(1420)$ to $\eta(1440)$ in the $K\bar{K}\pi$ channel is about
$17.3\%$.

\begin{table}[htbp]
  \centering
  \caption{Couplings extracted from the angular distribution analysis.
  }\label{tab:2}
  \begin{tabular}{cl}
    \hline\hline
    $g_{\eta(1440) K^* K}$   & 3.638\\
    $g_{J/\psi \gamma \eta(1440)}(\textrm{GeV}^{-1})$ & $(1.59\pm0.32)\times10^{-3}$\\
    $g_{f_1K^*K}(\textrm{GeV})$ &  $2.282\pm0.054$ \\
    $g_1$ & $(4.4\pm2.7)\times10^{-4}$ \\
    $g_2/g_1(\textrm{GeV}^{-2})$ & $-0.179_{-0.219}^{+0.403}$ or $0.970_{-0.403}^{+0.219}$ \\
    $\frac{\Gamma(J/\psi \to \gamma f_1 \to \gamma K\bar{K}\pi)}{\Gamma(J/\psi \to \gamma \eta(1440) \to \gamma K\bar{K}\pi)}$ &
    $(17.3\pm23.4)\%$ or $(17.1\pm23.1)\%$ \\
    \hline\hline
  \end{tabular}
\end{table}

\begin{figure}[htbp]
  \includegraphics[scale=1]{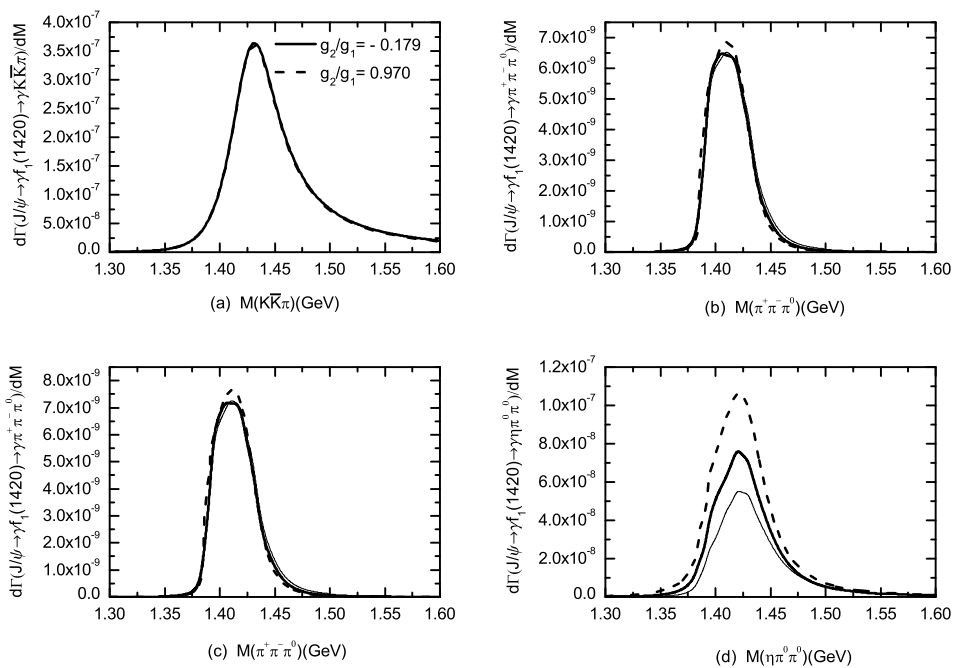}\\
\caption{Predictions for the spectra $\ud \Gamma(J/\psi \to \gamma
f_1(1420) \to \gamma ABC )/\ud \sqrt{s}$. Figure (a) is for the
$K\bar{K}\pi$ channel where the solid and dashed line denote results
with different values of $g_2/g_1$. Figures (b) and (c) are for the
$\pi^+\pi^-\pi^0$ channel with $g_{f_0 KK}$ and $g_{f_0 \pi\pi}$
determined by the BES and KLOE data, respectively, and (d) for the
$\eta\pi^0\pi^0$ channel. The thick and thin solid lines in Figures
(b-d) correspond to the results with $\Lambda=1.0 $ and $0.5 \
\textrm{GeV}$, respectively, while the dashed lines denote the
results by the {\it LoopTools} calculation without form factor.
  }
  \label{fig:f1420}
\end{figure}

With these couplings, we can predict the corresponding spectra and
ratios for $J/\psi\to \gamma f_1(1420)$ with $f_1(1420)\to
K\bar{K}\pi$, $\pi^+\pi^-\pi^0$ and $\eta\pi^0\pi^0$ as shown in
Fig.~\ref{fig:f1420} and Table~\ref{tab:3}. The results obtained by
those two values of $g_2/g_1$ are almost identical as demonstrated
in Fig.~\ref{fig:f1420}(a) for the $K\bar{K}\pi$ channel. So, in
other channels we only show the results with $g_2/g_1=-0.179$. The
main features of spectra and ratios are similar to those of
$\eta(1440)$.

In the $\pi^+\pi^-\pi^0$ channel, the results of KLOE are larger
than those of BESIII by a factor of about $1.11$ due to the
difference of $g_{fK\bar{K}}$ and $g_{f\pi\pi}$ extracted from these
two experiments as mentioned earlier. Meanwhile, it shows that the
partial width (or branching ratio) is insensitive to the form factor
cut-off energy. This feature has been discussed earlier and it is
because that the model uncertainties will be largely constrained by
the cancellation between the charged and neutral loop amplitudes.
Nevertheless, the dominant contributions to the isospin-violating
decays are from the TS kinematics where the form factor effects are
rather small.

In the $\eta\pi^0\pi^0$ channel, the results are sensitive to the
integration methods and cut-off energies due to the contributions
from the dispersive part in the loop integrals. When varying the
cut-off $\Lambda$ from $1.0 \ \textrm{GeV}$ to $0.5 \ \textrm{GeV}$,
the results change about $31\%$.

\begin{table}[htbp]
  \centering
\caption{ The extracted results for $f_1(1420)$, where
$\Gamma_{ABC}\equiv \Gamma(J/\psi \to \gamma f_1(1420) \to \gamma
ABC)$ and $R_{ABC}\equiv \Gamma_{ABC}/\Gamma_{K\bar{K}\pi}$.
  }\label{tab:3}
  \begin{tabular}{cccccccc}
    \hline\hline
    \multicolumn{2}{c}{coupling}                       &  \multicolumn{2}{c}{$g_2/g_1=-0.179 \ {\textrm{GeV}}^{-2}$}& \multicolumn{2}{c}{$g_2/g_1=0.970 \ {\textrm{GeV}}^{-2}$} \\ \hline
    \multicolumn{2}{c}{channel}                         & $\Gamma(\textrm{keV})$ & $R$         &  $\Gamma(\textrm{keV})$ & $R$     \\  \hline
    \multicolumn{2}{c}{$K\bar{K}\pi$}                   & $2.67\times10^{-2}$    & $ 1$        & $2.63\times10^{-2}$     & $ 1$ \\ \hline
                          &LoopTool                     & $3.18\times10^{-4}$    & $ 1.19\%$   & $3.18\times10^{-4}$     & $ 1.21\%$ \\
    $\pi^+\pi^-\pi^0$(BES)&$\Lambda=1.0 \ \textrm{GeV}$ & $3.06\times10^{-4}$    & $ 1.15\%$   & $3.06\times10^{-4}$     & $ 1.16\%$ \\
                          &$\Lambda=0.5 \ \textrm{GeV}$ & $3.11\times10^{-4}$    & $ 1.17\%$   & $3.11\times10^{-4}$     & $ 1.18\%$ \\ \hline
                          &LoopTool                     & $3.54\times10^{-4}$    & $ 1.33\%$   & $3.54\times10^{-4}$     & $ 1.34\%$ \\
   $\pi^+\pi^-\pi^0$(KLOE)&$\Lambda=1.0 \ \textrm{GeV}$ & $3.40\times10^{-4}$    & $ 1.27\%$   & $3.89\times10^{-4}$     & $ 1.29\%$ \\
                          &$\Lambda=0.5 \ \textrm{GeV}$ & $3.45\times10^{-4}$    & $ 1.29\%$   & $3.45\times10^{-4}$     & $ 1.31\%$ \\ \hline
                          &LoopTool                     & $6.78\times10^{-3}$    & $ 25.4\%$   & $6.74\times10^{-3}$     & $ 25.6\%$ \\
    $\eta\pi^0\pi^0$      &$\Lambda=1.0 \ \textrm{GeV}$ & $4.68\times10^{-3}$    & $ 17.5\%$   & $4.68\times10^{-3}$     & $ 17.8\%$ \\
                          &$\Lambda=0.5 \ \textrm{GeV}$ & $3.24\times10^{-3}$    & $ 12.1\%$   & $3.24\times10^{-3}$     & $ 12.3\%$ \\
    \hline\hline
  \end{tabular}
\end{table}

As we expect that the TSM appears to be more significant in
$f_1(1420)$ than in $\eta(1440)$ since the coupling $f_1(1420)\to
K^*\bar{K}+c.c.$ is in a relative $S$ wave, while $\eta(1440)\to
K^*\bar{K}+c.c.$ is in a $P$ wave. Taking the results with
$\Lambda=1.0 \ \textrm{GeV}$ as an example, the ratio of
$\pi^+\pi^-\pi^0$ to $K\bar{K}\pi$ is $1.27\%$ in $f_1(1420)$, while
in $\eta(1440)$ the ratio is $0.762\%$. The ratio of
$\eta\pi^0\pi^0$ to $K\bar{K}\pi$ is $17.5\%$ in $f_1(1420)$, while
in $\eta(1440)$ the ratio is $6.61\%$. The contributions from the
$f_1(1420)$ also affects the peak position as demonstrated in the
next Subsection.

\subsection{Invariant mass spectra including both $\eta(1440)$ and $f_1(1420)$}

\begin{figure}[htbp]
  \includegraphics[scale=1]{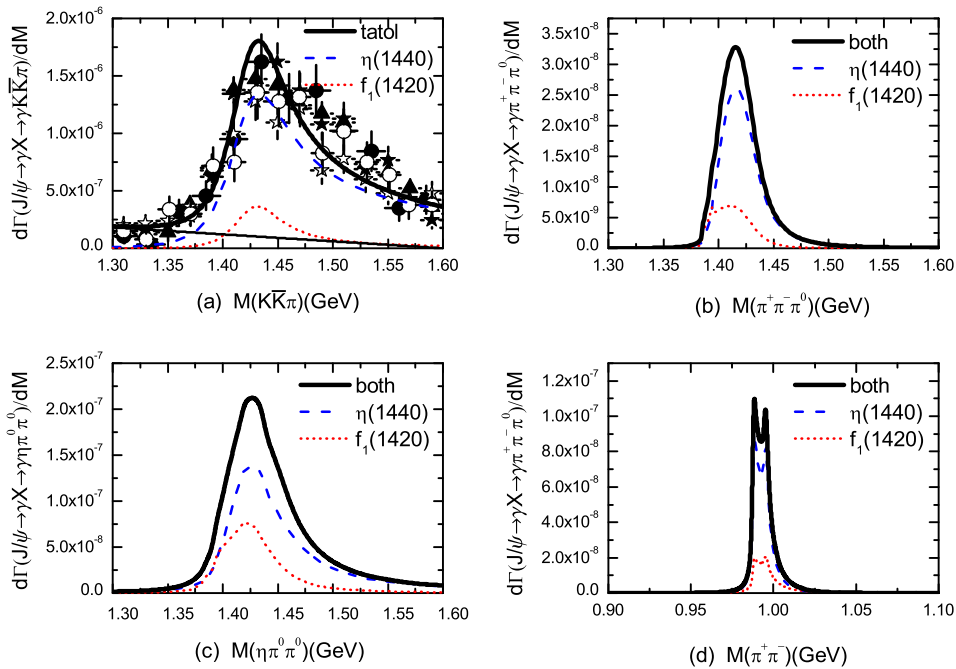}\\
\caption{The spectra $\ud \Gamma(J/\psi \to \gamma X \to \gamma ABC
)/\ud \sqrt{s}$ including both $\eta(1440)$ and $f_1(1420)$. In the
$K\bar{K}\pi$ channel, we also show the experimental data, i.e. the
solid triangles, solid circles, hollow circles, solid pentacles and
hollow pentacles from MARK III($K_S^0 K^\pm
\pi^\mp$)~\cite{Bai:1990hs}, BES($K_S^0 K^\pm
\pi^\mp$)~\cite{Bai:2000ss}, BES($K^\pm K^\mp
\pi^0$)~\cite{Bai:1998eg}, DM2($K_S^0 K^\pm
\pi^\mp$)~\cite{Augustin:1989zf}, and
DM2($K^+K^-\pi^0$)~\cite{Augustin:1989zf}, respectively. The thin
line in Fig. (a) is the background. In the $\pi^+\pi^-\pi^0$ and
$\pi^+\pi^-$ channels, we show the results given by the {\it
LoopTools} calculation with $g_{f_0 KK}$ and $g_{f_0 \pi\pi}$
determined by BES. In the $\eta\pi^0\pi^0$ channel, we choose the
results with $\Lambda=1.0 \ \textrm{GeV}$.
  }
  \label{fig:both}
\end{figure}

With the parameters fixed as the above, we compare the spectra and
ratios with experiment ~\cite{BESIII:2012aa} in Fig.~\ref{fig:both}
and Table~\ref{tab:4} where both  $\eta(1440)$ and $f_1(1420)$ are
included. The main features are consistent with
Ref.~\cite{Wu:2011yx} where only $\eta(1440)$ was considered.

It shows that the contribution of $f_1(1420)$ is much smaller than
that of $\eta(1440)$ in $J/\psi\to \gamma \eta(1440)/f_1(1420) \to
\gamma K\bar{K}\pi$. However, one should be reminded that this is
largely due to the suppressed coupling for $J/\psi\to\gamma
f_1(1420)$. In contrast, the contribution from the $f_1(1420)$ is
relatively enhanced in the $\eta\pi^0\pi^0$ channel than in
$K\bar{K}\pi$ because of the TSM. The most interesting scenario is
that the lineshapes of the invariant mass spectra for the
$K\bar{K}\pi$, $\eta\pi^0\pi^0$ and $\pi^0\pi^+\pi^-$ decays are
very different from each other due to the presence of the TSM in the
last two processes. Also, the interferences of the TSM have led to
the shifts of peak positions in those three channels which describe
the experimental data consistently. Such a phenomenon retains even
with contributions from $\eta(1440)$ exclusively as found in
Ref.~\cite{Wu:2011yx}.

For the isospin-violating channel of $\eta(1440)/f_1(1420)\to 3\pi$,
the observation of the narrow $f_0(980)$ in the $\pi\pi$ spectrum
can be regarded as a signature of the TSM. As being shown in
Ref.~\cite{Wu:2011yx}, the narrow peak is located between the
charged and neutral $K\bar{K}$ thresholds as a residual contribution
due to the isospin violation. The mass difference between the
charged and neutral kaons gives rise to the nonvanishing amplitudes
between the $K^+K^-$ and $K^0\bar{K}^0$ thresholds which has been a
crucial mechanism for the $a_0(980)$ and $f_0(980)$ mixing. Beyond
this scenario, what we show here and in Ref.~\cite{Wu:2011yx} is
that the TSM can further dominantly enhance the $f_0(980)$
production in $\eta(1440)/f_1(1420)\to 3\pi$ which eventually
explains the anomalously large isospin violations.

As listed in Table~\ref{tab:4}, we can see that the ratios of
$\pi^+\pi^-\pi^0$ and $\eta\pi^0\pi^0$ to $K\bar{K}\pi$ agree well
with experiment. Meanwhile, one also notices that the relative
contributions from $f_1(1420)$ to $\eta(1440)$ are quite different
in different channels as we discussed in Fig.~\ref{fig:both}.
Namely, the relative strength of $f_1(1420)$ to $\eta(1440)$ turns
out to be more significant in the $\eta\pi^0\pi^0$ channel than in
the $K\bar{K}\pi$. It is because the $S$-wave coupling of $K^*K$ to
$f_1(1420)$ would allow a relatively enhanced contributions from the
TSM in $J/\psi\to\gamma f_1(1420)\to \gamma \eta\pi^0\pi^0$ than
$\eta(1440)$.

\begin{table}[htbp]
  \centering
\caption{The combined results for $R_{ABC}=\Gamma(J/\psi \to \gamma
X \to \gamma ABC)/\Gamma(J/\psi \to \gamma X \to \gamma
K\bar{K}\pi)$ including both $\eta(1440)$ and $f_1(1420)$. The
experimental
data~\cite{Bai:2000ss,Bai:1999tg,Li:2011ve,BESIII:2012aa} are also
listed. For $f_1(1420)$ we adopt $g_2/g_1=-0.179$. }\label{tab:4}
  \begin{tabular}{cccc}
    \hline\hline
    \multicolumn{4}{c}{R} \\ \hline
    \multicolumn{2}{c}{channel} & Theory & Expt. \\ \hline
    \multicolumn{2}{c}{$K\bar{K}\pi$}                  & 1    & 1 \\ \hline
                           & LoopTool                  & $0.781\%$ &  \\
    $\pi^+\pi^-\pi^0$(BES) & $\Lambda=1.0 \ \textrm{GeV}$ & $0.746\%$ & $(0.90\pm0.39)\%$ \\
                           & $\Lambda=0.5 \ \textrm{GeV}$ & $0.752\%$ &  \\ \hline
                           & LoopTool                  & $0.878\%$ &  \\
    $\pi^+\pi^-\pi^0$(KLOE)& $\Lambda=1.0 \ \textrm{GeV}$ & $0.837\%$ & $(0.90\pm0.39)\%$ \\
                           & $\Lambda=0.5 \ \textrm{GeV}$ & $0.843\%$ &  \\ \hline
                           & LoopTool                  & $23.7\%$ &  \\
    $\eta\pi^0\pi^0$       & $\Lambda=1.0 \ \textrm{GeV}$ & $8.22\%$ & $(7.8\pm4.6)\%$ \\
                           & $\Lambda=0.5 \ \textrm{GeV}$ & $5.28\%$ &  \\
    \hline\hline
  \end{tabular}
\end{table}

In Table \ref{tab:5} we present the $\eta(1440)$ and $f_1(1420)$
peak positions extracted in those three decay channels. Due to the
contributions from the TSM, the peak positions are shifted
differently. It shows that the exclusive results for $\eta(1440)$
and $f_1(1420)$ respectively or the results with their combined
contributions have a similar feature. Namely, the largest peak mass
can be seen in the $K\bar{K}\pi$ channel, while the smallest one is
the $3\pi$ channel. This qualitative pattern fits well the
experimental observations in these three decay channels.

Because of the TSM, the peak positions in both $\eta\pi\pi$ and
$3\pi$ channels would move towards the $K^*\bar{K}+c.c.$ threshold
which is about $1.39 \ \textrm{GeV}$. The more significant the TSM
contribution is, the larger the peak position shift would be. As a
result of the TSM dominance in the $\pi^+\pi^-\pi^0$ channel, the
peak position observed in the $\pi^+\pi^-\pi^0$ channel has a lower
value than that in the $\eta\pi^0\pi^0$ channel. The importance of
the TSM suggests that a partial wave analysis including the TSM is
necessary. Such a mechanism may also have significant interferences
with the background. As a consequence, it will lead to different
lineshapes for the $\eta(1440)$ in different production channels.
For instance, the peak position of the $\eta(1440)$ in $J/\psi\to
\gamma \eta \pi\pi$ is slightly different from that in $J/\psi\to
\omega\eta\pi\pi$~\cite{Ablikim:2011pu}. Note that the results of
Ref.~\cite{Ablikim:2011pu} are given by simple Breit-Wigner fit
instead of partial wave analysis. Further detailed analysis of this
channel using partial wave analysis should include the TSM as an
important underlying dynamics in order to extract the correct pole
position for the $\eta(1440)$.

With $m_{\eta(1440)}=1.42 \ \textrm{GeV}$ and $m_{f_1(1429)}=1.4264
\ \textrm{GeV}$,  we find that the peak position shifts in the
$f_1(1420)$ decays are larger than in $\eta(1440)$. The reason again
is because of the relative $S$-wave coupling for  $f_1(1420)\to
K^*\bar{K}+c.c.$ When combining $\eta(1440)$ and $f_1(1420)$
together, the largest peak position shift that we can achieve is
about $24 \ \textrm{MeV}$, which supports our one-state assumption.
Namely, the $\eta(1405)$ and $\eta(1475)$ may be just one state in
different channels.

\begin{table}[htbp]
  \centering
  \caption{Peak positions in different channels.}\label{tab:5}
  \begin{tabular}{ccccc}
    \hline\hline
    peak position (GeV)       & $K\bar{K}\pi$  & $\pi^+\pi^-\pi^0$  & $\eta\pi^0\pi^0$ \\ \hline
    $\eta(1440)$              & 1.433          & 1.416              & 1.426 \\
    $f_1(1420)$               & 1.431          & 1.411              & 1.422 \\
    $\eta(1440)+f_1(1420)$    & 1.432          & 1.415              & 1.425 \\
    \hline\hline
  \end{tabular}
\end{table}

\subsection{Radiative decays of $\eta(1405/1475)$}

Our proposal that $\eta(1405)$ and $\eta(1475)$ are the same state
would have an explicit consequence in the description of the
radiative decays of $\eta(1405/1475)\to \gamma V$, where $V$ stands
for the light vector mesons $\phi$, $\rho^0$ and $\omega$. In the
one-state assumption, $\eta(1440)$ would be the SU(3) flavor partner
of $\eta(1295)$ as the first radial excitation states of $\eta$ and
$\eta'$. In Ref.~\cite{Klempt:2007cp}, it was commented that by
assigning the $\eta(1475)$ to the SU(3) partner of $\eta(1295)$ as
the $s\bar{s}$ dominant state would not be able to explain why the
observed branching ratios $BR(\eta(1475)\to \gamma \rho^0))$ is
larger than  $BR(\eta(1475)\to \gamma \phi))$. Also, it was
commented that the observation that the much stronger production
rate of $J/\psi\to \gamma\eta(1405/1475)$ than $J/\psi\to
\gamma\eta(1295)$ seemed not be obvious taking into account the
above question. However, in this Subsection, we shall show that the
experimental observations can be self-consistently understood by
treating $\eta(1440)$ and $\eta(1295)$ as the SU(3) flavor partners.
This can be explicitly demonstrated as the following:

By assigning $\eta(1295)$ and $\eta(1440)$ as the first radial
excitation of $\eta$ and $\eta'$, we can organize them as the
following mixtures between $n\bar{n}\equiv
(u\bar{u}+d\bar{d})/\sqrt{2}$ and $s\bar{s}$:
\begin{eqnarray}
\eta(1295)&= &\cos\alpha n\bar{n}-\sin\alpha s\bar{s} \nonumber\\
\eta(1440) &= &\sin\alpha n\bar{n} +\cos\alpha s\bar{s} \ ,
\end{eqnarray}
where $\alpha$ is the mixing angle.

In the $J/\psi$ radiative decays, it is a good approximation that
the photon is radiated by the charm (anti-)quark, and the light
$q\bar{q}$ of $0^{-+}$ is produced by the gluon radiation. By
defining the production strength for the $q\bar{q}$ of $0^{-+}$ as
the following:
\begin{equation}
g_0\equiv \langle q\bar{q}|\hat{H}|J/\psi,\gamma\rangle \ ,
\end{equation}
one can express the production amplitudes for $\eta(1295)$ and
$\eta(1440)$ as
\begin{eqnarray}
{\cal M}(\eta(1295))&=& (\sqrt{2}\cos\alpha-R\sin\alpha)g_0 \ ,
\nonumber\\
{\cal M}(\eta(1440)) &=& (\sqrt{2}\sin\alpha+R\cos\alpha) g_0 \ ,
\end{eqnarray}
where $R\equiv \langle s\bar{s}|\hat{H}|J/\psi,\gamma\rangle/g_0$ is
an SU(3) flavor symmetry breaking factor, and one can simply set it
to be unity as a leading approximation. It can be easily seen that a
proper value for the mixing angle $\alpha$ in the first quadrant
would lead to a much suppressed b.r. for $J/\psi\to
\gamma\eta(1295)$ than for $J/\psi\to \gamma\eta(1440)$. The value
of $\alpha$ can be determined by the b.r.s measured for these two
channels. For instance, if one requires that $B.R.(J/\psi\to
\gamma\eta(1440))/B.R.(J/\psi\to \gamma\eta(1295))\simeq 10$,
namely, the production of $\eta(1440)$ is about one order of
magnitude larger than $\eta(1295)$, one would have
\begin{equation}
\frac{B.R.(J/\psi\to \gamma\eta(1440))}{B.R.(J/\psi\to
\gamma\eta(1295))}=
\left(\frac{q_{\eta(1440)}}{q_{\eta(1295)}}\right)^3
\left(\frac{\sqrt{2}\sin\alpha+R\cos\alpha}{\sqrt{2}\cos\alpha-R\sin\alpha}\right)^2\simeq
10 \ ,
\end{equation}
where $q_{\eta(1440)}$ and $q_{\eta(1295)}$ are three momenta of the
pseudoscalars in the $J/\psi$ rest frame, respectively. with
$R\equiv 1$, one has $\alpha\simeq 38^\circ$. Such a mixing scenario
will have explicit predictions for the radiative decays of
$\eta(1440)\to \phi\gamma$, $\rho^0\gamma$ and $\omega\gamma$.

Since $\phi$ and $\omega$ are nearly ideally mixed to each other and
$\rho^0$ has isospin-1, we adopt the flavor wavefunctions,
$\phi=s\bar{s}$ and $\omega= n\bar{n}$, and
$\rho^0=(u\bar{u}-d\bar{d})/\sqrt{2}$. The $\eta(1440)$ radiative
decays are via M1 transitions where the quark spin will be flipped
by the magnetic interaction. A standard operator in the quark model
can be written as
\begin{equation}
\hat{H}_{em}\equiv \langle \phi_A\chi_S|\sum^2_i
e_i\mu_i\overrightarrow{\sigma}_i\cdot\overrightarrow{\epsilon}_\gamma|\phi_S\chi_A\rangle
\ ,
\end{equation}
where $\mu_i\equiv e/2m_i$ is the magnetic moment of the $i$th
quark, and $|\phi_S\chi_A\rangle$ and $|\phi_A\chi_S\rangle$ are the
flavor-spin wavefunctions for $\eta(1440)$ and vector meson,
respectively. The subscriptions $S$ and $A$ means that the
corresponding wavefunctions are symmetric or anti-symmetric under
the exchange of the first and second quark (anti-quark). The flavor
and spin wavefunctions are defined in a standard way as the
following:
\begin{eqnarray}
\phi_S(s\bar{s})&\equiv & (s\bar{s}+\bar{s}s)/\sqrt{2} \ ,
\nonumber\\
\phi_S(n\bar{n})&\equiv & (n\bar{n}+\bar{n}n)/\sqrt{2} \ ,
\nonumber\\
\chi_A &\equiv & (\uparrow\downarrow-\downarrow\uparrow)/\sqrt{2}
\end{eqnarray}
for the pseudoscalar state, and
\begin{eqnarray}
\phi_A(\phi)&\equiv & (s\bar{s}-\bar{s}s)/\sqrt{2} \ ,
\nonumber\\
\phi_A(\rho^0)&\equiv & ((u\bar{u}-\bar{u}u)-(d\bar{d}-\bar{d}d))/2
\ ,
\nonumber\\
\phi_A(\omega)&\equiv & ((u\bar{u}-\bar{u}u)+(d\bar{d}-\bar{d}d))/2
\ ,
\nonumber\\
\chi_S &\equiv & \uparrow\uparrow, \ \downarrow\downarrow, \
(\uparrow\downarrow+\downarrow\uparrow)/\sqrt{2} \ ,
\end{eqnarray}
for the vectors.

One can easily work out the flavor-spin couplings for those three
channels as follows:
\begin{eqnarray}
h_{\phi\gamma}&=& -\frac{e}{3m_s}\cos\alpha \ ,\nonumber\\
h_{\rho^0\gamma}&=& \frac{e}{2m_q}\sin\alpha \ ,\nonumber\\
h_{\omega\gamma}&=& \frac{e}{6m_q}\sin\alpha \ ,
\end{eqnarray}
where $m_q=m_u=m_d$ and  $m_s\simeq 5m_q/3$. Apart from the spacial
form factor and phase space factor in a $P$ wave, the b.r. fraction
among these decay channels are then
\begin{equation}
B.R.(\gamma\phi):B.R.(\gamma\rho^0):B.R.(\gamma\omega)\simeq
\frac{\cos^2\alpha}{25}:\frac{\sin^2\alpha}{4}:\frac{\sin^2\alpha}{36}
\ .
\end{equation}
For a proper value of $\alpha$ in the first quadrant, the decay of
$\eta(1440)\to \gamma\rho^0$ would be dominant. To be consistent
with the production of $\eta(1440)$ and $\eta(1295)$ in the $J/\psi$
radiative decays, i.e. $\alpha\simeq 38^\circ$, one obtains
$B.R.(\gamma\phi):B.R.(\gamma\rho^0):B.R.(\gamma\omega)\simeq  1:
3.8: 0.42$.

In brief, given a proper mixing angle between the $\eta(1295)$ and
$\eta(1440)$ as the first radial excitation states of $\eta$ and
$\eta'$, the theoretical interpretation of the $\eta(1405/1475)$ as
a single state of $\eta(1440)$ does not obviously conflict with the
so far available experimental data at all. The misunderstanding that
the branching ratio of $\eta(1475)\to \gamma\phi$ should be larger
than that of $\eta(1475)\to \gamma\rho^0$ if $\eta(1475)$ is the
higher mass partner of $\eta(1295)$ is not necessary at all due to
the suppression of the quark masses in the M1 transition. This
point, unfortunately, has not been realized in earlier analyses.

\section{Summary}

In summary,  we have made a systematic analysis of the correlated
processes $J/\psi\to \gamma \eta(1440)/f_1(1420)$ with
$\eta(1440)/f_1(1420)\to K\bar{K}\pi$, $\eta\pi\pi$ and $3\pi$,
where the role played by the TSM is clarified. Our combined analysis
including $\eta(1440)$ and $f_1(1420)$ agrees well with the
experiment data, and provides an overall description of the
processes $J/\psi\to \gamma X$ with $X\to K\bar{K}\pi$,
$\eta\pi^0\pi^0$, and $\pi^+\pi^-\pi^0$. In particular, we show that
the inclusion of the $f_1(1420)$ can improve the description of the
$f_0(980)\pi^0$ angular distribution significantly, although the
contribution from $f_1(1420)$ is much smaller than $\eta(1440)$. By
fitting the BESIII data for $J/\psi \to \gamma X \to \gamma
f_0(980)\pi^0$, we extract the coupling parameters of $f_1(1420)$.
It allows us to estimate that the ratio of $f_1(1420)$ to
$\eta(1440)$ in the $K\bar{K}\pi$ channel is about $17.3\%$. This
does not change the results of the previous work~\cite{Wu:2011yx} in
which we assumed that $\eta(1440)$ was the only contributing state
as treated by the BESIII. We also show that $f_1(1420)$ can
contribute some percentages to the narrow peak of $f_0(980)\to
\pi\pi$ via the TSM.

We emphasize that the dynamic feature of the TSM can be recognized
by the strong narrow peak observed in the $3\pi$ channel with the
anomalously large isospin violations. Moreover, it leads to the
obvious peak position shifts for the same $\eta(1440)$ or
$f_1(1420)$ state in different decay channels, which may suggest
that the $\eta(1405)$ and $\eta(1475)$ are actually the same state.
So far, such a one-state prescription seems not to have a conflict
with existing experimental data. This may shed a light on the
long-standing puzzling question on the nature of $\eta(1405)$ and
$\eta(1475)$ in the literature.

\section{acknowledgments}

Useful discussions with X.-H. Liu, X.-Y. Shen, and Z. Wu are
acknowledged. This work is supported, in part, by the National
Natural Science Foundation of China (Grant Nos. 11035006 and
11121092), DFG and NSFC (CRC 110), Chinese Academy of Sciences
(KJCX2-EW-N01), Ministry of Science and Technology of China
(2009CB825200), and U.S. Department of Energy (Contract No.
DE-AC02-06CH11357).

\bibliography{reference}

\end{document}